MASTER THESIS

Julien **TARAN** - ROB 2023

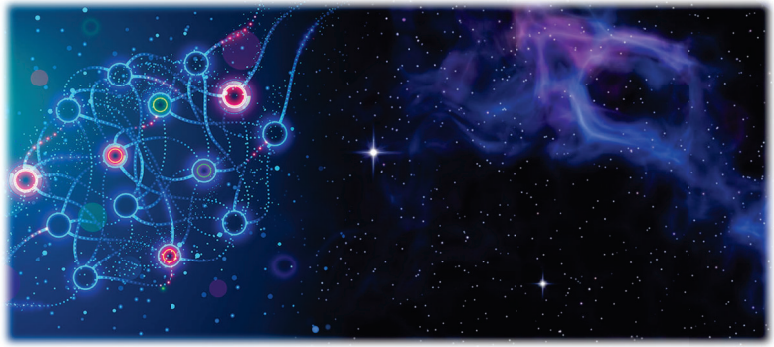

# Convolutional Neural Network For Lyman Break Galaxies Classification And Redshift Regression in DESI (Dark Energy Spectroscopic Instrument)

MARCH - JULY 2023

Supervisor : **Christophe YECHE** (CEA)
Tutor : **Luc JAULIN** (ENSTA Bretagne)

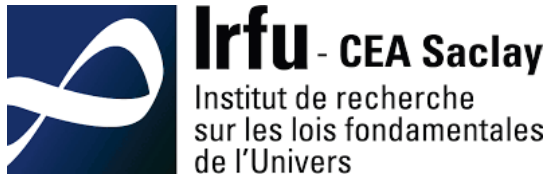
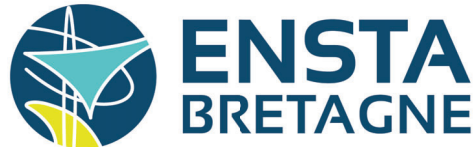

Particle Physics Division - DPHP
Institute of Research into the Fundamental Laws of the Universe - IRFU
French Alternative Energies and Atomic Energy Commission - CEA
Saclay, France

# Acknowledgements

Before beginning this report, I'd like to thank a number of people who helped me a lot during this final year internship.

First of all, I would like to extend my special thanks to Christophe Yeche, who proposed this internship to me, then supervised me throughout. He gave me excellent guidance on the ideas to explore, and kept a close eye on my progress, while remaining flexible and attentive to the problems I encountered.

Thanks to Luc Jaulin, my supervisor at ENSTA Bretagne, for his support and flexibility during this internship. More generally, I'd like to thank the teaching staff at ENSTA Bretagne for the courses given during the last two years of my specialization, which gave me the skills I needed to carry out this internship.

I'd also like to thank the entire cosmology team, in particular PhD students Edmond Chaussidon, Antoine Rocher, Alexandre Huchet, Mathilde Pinon and Marie Lynn Abdul Karim, who very quickly integrated me into the team, where I quickly felt at ease. It was very interesting for me to be able to talk to passionate people who taught me a lot about the universe and its mysteries.

# Abstract

DESI is a ground-breaking international project to observe more than 40 million quasars and galaxies over a 5-year period to create a 3D map of the sky. This map will enable us to probe multiple aspects of cosmology, from dark energy to neutrino mass. We're focusing here on one type of object observed by DESI, the Lyman Break Galaxies (LBGs). The aim is to use their spectra to determine whether they are indeed LBGs, and if so, to determine their distance from the Earth using a phenomenon called redshift. This will enable us to place these galaxies on the DESI 3D map.

The aim is therefore to develop a convolutional neural network (CNN) inspired by QuasarNET (See [7]) performing simultaneously a classification (LBG type or not) and regression task (determine the redshift of the LBGs). Initially, data augmentation techniques such as shifting the spectra in wavelenghts, adding noise to the spectra, or adding synthetic spectra were used to increase the model training dataset from 3,019 data to over 66,000. In a second phase, modifications to the QuasarNET architecture, notably through transfer learning and hyperparameter tuning, with Bayesian Optimization, boosted model performance.

Gains of up to 26% were achieved on the Purity/Efficiency curve, which is used to evaluate model performance, particularly in areas with interesting redshifts, at low (around 2) and high (around 4) redshifts. The best model obtained an average score of 94%, compared with 75% for the initial model.

# Contents

# Acronymes

**CEA :** Centre à l'Energie Atomique et aux énergies alternatives (French alternative energies and atomic energy commission)
**CNN :** Convolutional Neural Network
**DESI :** Dark Energy Spectroscopic Instrument
**ELG :** Emission-Line Galaxy
**IRFU :** Institut de Recherche sur les lois Fondamentales de l'Univers
**LAE :** Lyman-$\alpha$ Emitters
**LBG :** Lyman Break Galaxy
**Ly-$\alpha$ :** Lyman-$\alpha$
**QSO :** Quasi-Stellar Object
**RR :** Redrock
**SNR :** Signal-to-Noise Ratio
**TL :** Transfer Learning

# Chapter 1

# Introduction

## 1.1 Presentation of the CEA and the IRFU

### 1.1.1 The CEA

The CEA (The French Alternative Energies and Atomic Energy Commission) in Saclay, France is a world-renowned prestigious research facility in the South of Paris. It occupies a central position in the field of science and technology in France.

The work of CEA Saclay is very diverse and includes significant contributions in the areas of nuclear fusion, advanced nuclear reactors, materials and nanoscience, modeling and simulation, nuclear medicine, renewable energy, and many more.

The CEA was founded by General de Gaulle, who wanted to set up a nuclear research organization at the end of World War II. However, since the 2000s, CEA has expanded its activities into renewable energy (solar energy, fuel and hydrogen cells, biofuels) and new technologies (quantum computing, artificial intelligence, self-driving cars, robotics).

### 1.1.2 The IRFU

IRFU, the Institute for Research into the Fundamental Laws of the Universe, part of the CEA's Fundamental Research Division (DRF) and located on the Saclay campus, brings together three scientific disciplines: astrophysics, nuclear physics and particle physics, and all related technical expertise.

It is positioned to answer the main open questions in the understanding of the four fundamental interactions, on different scales, from the smallest (elementary constituents of matter, nuclear matter) to the largest (energy content and structures of the Universe).

In order to answer key questions concerning the elementary constituents of matter, the organization of nuclear matter, the structure and energy content of the Universe, Irfu is organized around scientific themes, to which are added technological themes linked to particle accelerators and superconducting magnets, detectors, microelectronics circuits and signal processing, as well as a theme linked to algorithmic developments for scientific computing.

The internship took place within the Cosmology team of the IRFU's Department of Particle Physics (DPhP). The core of DPhP's activities is the understanding of the energy content and dynamics of the Universe at the fundamental level, by acquiring the appropriate experimental means, in synergy with the other Irfu's departments, and in concert with the international scientific strategy (See [1] for more).

## 1.2 The DESI Program

DESI (Dark Energy Spectroscopic Instrument) began observing the sky on May 15, 2021, and will continue to do so for a total experiment of 5 years (See [2]). The first public data were published on June 13 2023, with over 2 million objects already observed. Installed in the Mayall Telescope at the Kitt Peak National Observatory, shown in Figure 1.1, it is part of an international collaboration in which the CEA is a key player, and the work of this internship is part of this collaboration.

The technology of this instrument is quite remarkable: 5,000 robotized fibers can simultaneously observe 5,000 celestial objects and plot their spectra using a bank of spectrographs powered by these fibers.

The aim is to observe a total of 40 million galaxies and quasars over 5 years, in order to produce a 3D map of the sky. The first two dimensions of this 3D map, corresponding to the position of celestial objects in the sky, are available thanks to the photometric surveys carried out by the DESI Legacy Imaging Survey (See [9]), whose aim is to photograph the sky to locate objects in the sky. The challenge is then to identify the objects of interest (target selection stage), which will then be observed by DESI's 5,000 fibers.

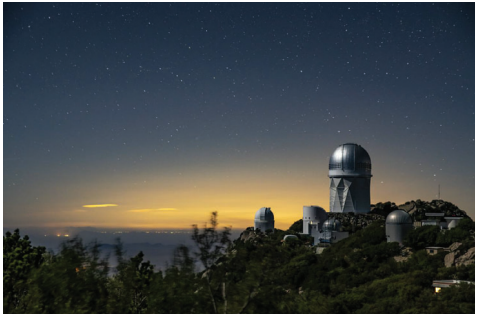

Figure 1.1: *A view of the Mayall Telescope. Credit : Lawrence Berkeley National Laboratory.*

The spectrographic readings from these 5,000 fibers will then be used to determine the final dimension of this 3D map: the distance of the objects from us. This distance is determined from the spectra of these objects, thanks to their redshift. This phenomenon is developed in the next section.

This 3D map of the sky is of enormous interest from a scientific point of view, and will enable us to probe multiple aspects of cosmology, from dark energy to neutrino mass.

The DESI collaboration provides access to the National Energy Research Scientific Computing Center (NERSC), operated by Lawrence Berkeley National Laboratory, and in particular to its latest supercomputer Perlmutter, the 5th fastest supercomputer in the world. This greatly facilitated the learning process on deep learning models with numerous parameters, carried out during the internship.

## 1.3 Introduction to the cosmological redshift

Redshift is a direct way of measuring the distance of objects from Earth.

The universe is expanding, and this expansion has the effect of stretching the light, because of a phenomenon called redshift, which is analogous to the Doppler effect in the case of distant galaxies. The longer a photon takes to travel through this expanding space before reaching the Earth, the longer its wavelength will increase as a result of this expansion. As a result, all wavelengths in a spectrum undergoing redshift are shifted to longer wavelengths. In the visible spectrum, wavelengths shift towards red, the color with the longest visible wavelengths, hence the name "redshift".

The idea is to identify emission or absorption lines of chemical elements at known wavelengths, and to pinpoint their wavelength shift. In this way, we can deduce the object's redshift and then infer its distance. This is how the 3rd dimension of the DESI's 3D map is infered.

Noted $z$, the relationship between the redshift, the observed wavelength and the wavelength in the rest frame, i.e. if the object was right next to the Earth, is given by

$$z = \frac{\lambda_{obs} - \lambda_{rf}}{\lambda_{rf}}$$

With $\lambda_{obs}$ the observed wavelength and $\lambda_{rf}$ the wavelength in the rest frame. See [6], Chapter 3 for more.

## 1.4   The Lyman Break Galaxies and their spectra

Lyman-break galaxies (LBG) are one of the tracers observed with interest by DESI.

The LBGs are star-forming galaxies distinguishable in that radiation with energies below the Lyman limit (at 912 Å) is absorbed almost entirely by the neutral gas surrounding the star-forming regions of the galaxies. In the rest frame of LBGs (when the LBG has a redshift of 0, as if it was right next to our planet in the universe), the spectrum is luminous at wavelengths above this limit, but gets fainter and fainter at shorter wavelengths - this phenomenon is called "Lyman Break" and is the explaination of the name "Lyman Break Galaxies".

Light with wavelengths below 912 Å is in the far ultraviolet range and is absorbed by Earth's atmosphere, but for very distant galaxies, the wavelength of light is greatly stretched due to the expansion of the Universe (as explained in Section 1.3), and thus is not absorbed anymore. The spectra of two LBGs are shown in Figure 1.2. LBGs can be recognized not only by the general shape of their spectrum, but also by the various absorption lines (we'll be focusing on 15 of these in the CNN presented in Section 2.2) and one emission line, the $Ly-\alpha$ line (See Appendix B.1 to see all the lines). This emission line is more or less pronounced, making the galaxy classification more or less obvious.

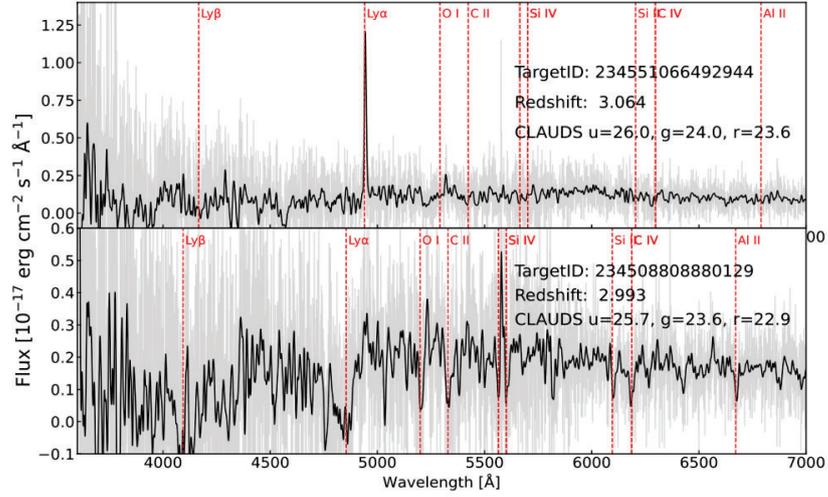

Figure 1.2: *A LBG spectra with a strong $Ly-\alpha$ emission line (upper plot) and a LBG spectra without $Ly-\alpha$ emission line (lower plot). Credit: Christophe Magneville.*

When this line is well pronounced, this type of galaxy is also called Lyman Alpha Emiters (LAE), and an example of the spectrum of these LBG can be seen in the upper plot of Figure 1.2 . Below, we can see a LBG that has no $Ly-\alpha$ emission line, making it much more complex to identify. Around 20% of LBGs have such a spectrum, without $Ly-\alpha$ emission line. See [3] for more.

# Chapter 2

# A Convolutional Neural Network for Classification and Regression

## 2.1 A 2-type task

In order for DESI's 5000 fibers to observe only LBGs, we need to be able to indicate the instrument where to look in the sky. To do this, the first stage, known as "Target Selection" (See [9]), aims to determine the positions of LBGs from photometric surveys so that DESI can dress their spectrum and then we can infer their redshifts.

The aim of the Convolutional Neural Network (CNN) is to ensure that the object is effectively a LBG, and then to determine its redshift, so there are 2 tasks to be carried out simultaneously.

### 2.1.1 A Classification task

Target selection of the LBGs is based on color-color analysis (See [4] fo the principle and [11] for the used analysis) using photometry. Figure 2.1 shows the target selection of LBGs based on COSMOS readings. The area delimited by the blue line is the selection zone. We can see that the target selection avoids the vast majority of stars (red dots) and other contaminants, but some targets are inevitably selected, notably a number of ELGs, with z values between 0.05 and 0.15 (left plot) and QSOs with redshifts greater than 2 (right plot).

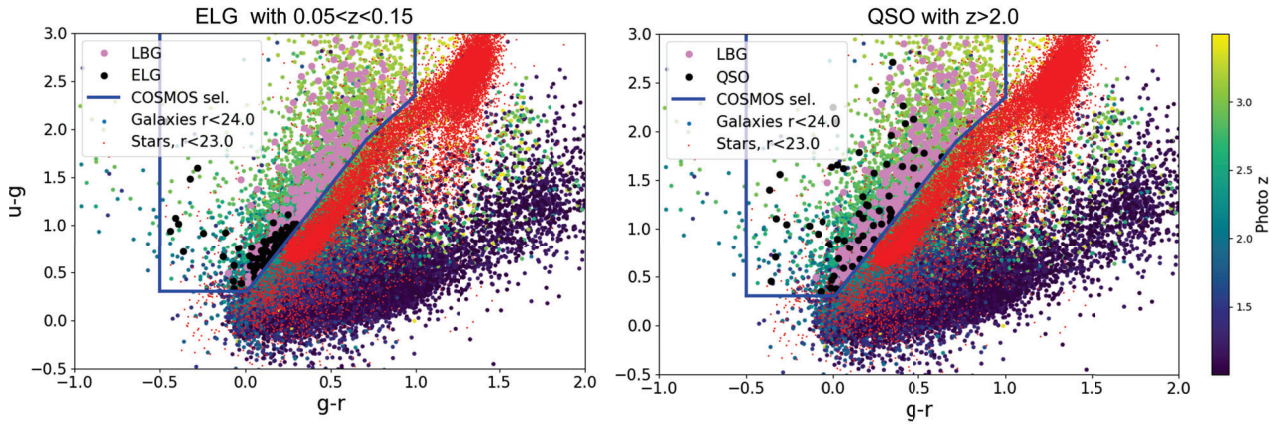

Figure 2.1: *The target selection of LBGs. Credit: Christophe Yeche.*

Thus, their spectra will be surveyed with DESI, in the same way as LBGs, as they cannot be distinguished at this stage.

The first objective of the CNN is therefore to separate the LBG spectra from the QSO and ELG spectra. This is a classification task, aimed at separating LBGs from contaminants (QSOs, ELGs).

### 2.1.2 A Regression task

In addition to distinguishing the nature of the object, we need to determine its redshift, and thus succeed in finding the position of characteristic lines in the spectrum (See Appendix B.1 for details on the lines). Once we know the position of a line, we can determine the galaxy's redshift, and thus deduce its distance from Earth.

The CNN's second mission is therefore to determine the position on the spectrum of the 16 characteristic lines of the LBGs.

## 2.2 QuasarNET and its architecture

### 2.2.1 The model's layers

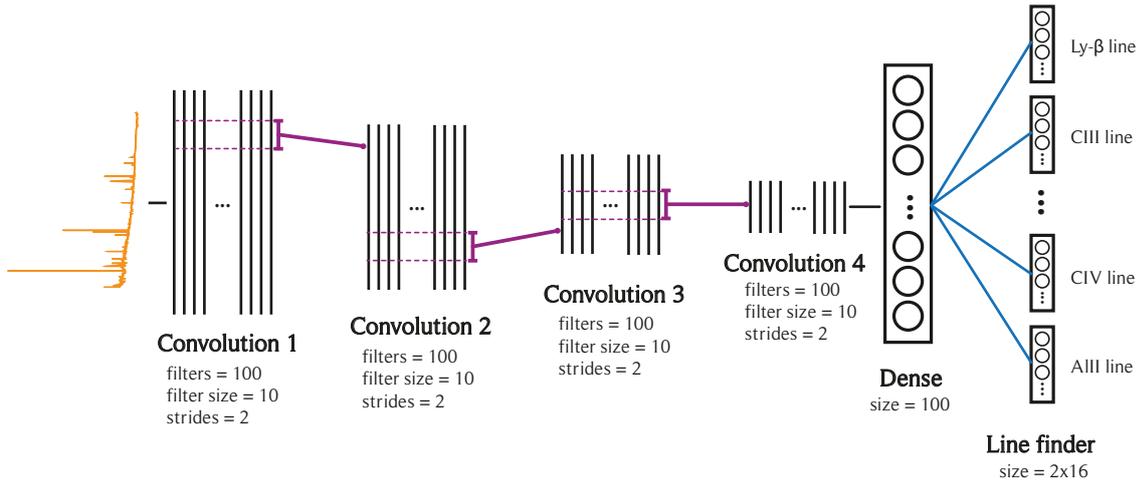

Figure 2.2: *Simplified architecture of QuasarNET CNN model.*

QuasarNET (See [7]) is a convolutional neural network with $p = 851,780$ trainable parameters. Its architecture is shown in Figure 2.2. It is inspired by architectures commonly used in Computer Vision, whose aim is to locate objects in images. Initially, four convolution layers are used to extract spectrum features. Due to the "black box" nature of neural networks, it is very difficult to get an idea of which features are being learned by the network. Each of the four layers is separated by a Batch Normalization layer (not shown in the figure, see Appendix D.1 for details), which reduces the activations to a mean of zero and a standard deviation of one, thus improving the stability and convergence speed of the network. Following this Batch-Normalization layer, a non-linear Rectified Linear Unit (ReLu) activation function (see Appendix D.2 for details on the ReLu) introduces the non-linearity required for the model to learn complex relationships between the data samples. Following these 4 convolution layers is a "Flatten" layer, which transforms the output of the convolution layers into a one-dimensional tensor, which is then connected to a Dense layer with 100 neurons ("Dense" layer on the diagram).

The most interesting and distinctive part of QuasarNET is its "Line Finder", the final stage of the network. The idea is to create a doublet of two groups of neurons for each absorption or emission line in the spectrum. Each of these two groups will be made up of $n_{boxes}$ neurons, each corresponding to one of the $n_{boxes}$ "sliced" regions of the spectrum. For our application to LBGs, we determined that 40 boxes was an effective parameter.

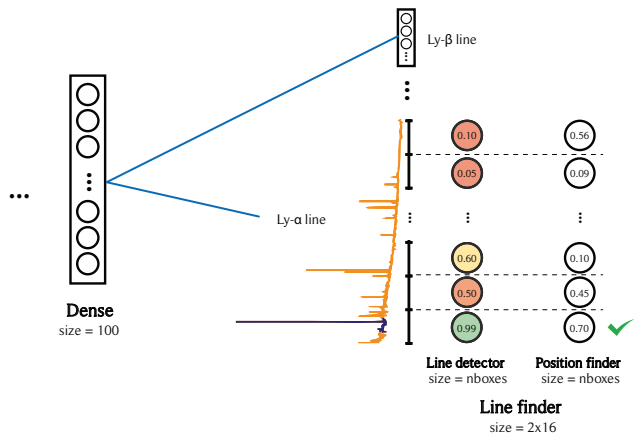

Figure 2.3: *Detailed architecture of the Line Finder.*

Figure 2.3 shows how this line finder works in detail. The role of the first group of neurons, called "Line Detector" in the figure, is to determine whether the line is included in this part of the spectrum. The closer the number is to 1 (green neuron in the figure), the greater the chance that the line is in the "box", i.e. the corresponding region of the spectrum. The second group of neurons, the position finder, determines the position of the line within the box. If the number is close to 0, the line will be close to the left edge of the interval, and if it's close to 1, to the right edge. On the figure, we can see that the CNN locates the $Ly - \alpha$ line at 0.99 confidence in the first box, with an exact position at 0.7 from the left edge of the interval.

Once these predictions have been made by QuasarNET, we only keep the value of the "Line detector" neuron with the highest value. This value is called the Confidence Level (CL or c_th). We also have the associated position, determined with the "Position finder". We do this with all the lines, so in the case of LBGs we have 16 pairs $(CL, Position)$. From these pairs $(CL, Position)$, we can determine whether it is indeed an LBG. To do this, we look to see if the CL of the best line detected (in reality, we take the CL of the 5th best line) is greater than a value we set, typically 0.95. If we have $CL > 0.95$, we consider that the CNN is fairly confident and that this is indeed a LBG. We can then determine the position of this line on the spectrum using the position finder value, and thus deduce the redshift of the spectrum. In this way, the desired classification and regression tasks are performed simultaneously.

### 2.2.2 The loss function

As mentioned above, the purpose of the CNN is therefore twofold, and this must be taken into account for the Loss Function, a very important building block in neural network training.

The principle of the loss function is as follows, in very simplified terms. The CNN is asked to predict results from a given input spectrum. If the results are very different from those expected, we want to have a high Loss value, and then "punish" the CNN, that is to say significantly correcting its weights. On the other hand, if the results are close to those expected, we want a low loss value, so we will "encourage" the CNN and will not change significantly its weights. In our case, this learning process uses Adam optimization, which is an extension of the stochastic gradient descent (SGD) algorithm. See [13] for more details.

On the one hand, we want to train the CNN to set the CL of all its lines close to 0 in the case of a contaminant (ELG or QSO), and close to 1 only in the case of an LBG and only in the right box. On the other hand, we want to train it to find the exact position of the line within the box, so that we can then deduce the redshift. The loss function used is the following one:

$$\mathcal{L} = \sum_{\ell} \left[ -\frac{1}{\sum_{i\alpha} Y_{i\ell\alpha}} \sum_{i\alpha} Y_{i\ell\alpha} \ln \hat{Y}_{i\ell\alpha} \right.$$
$$- \frac{1}{\sum_{i\alpha}(1 - Y_{i\ell\alpha})} \sum_{i\alpha} (1 - Y_{i\ell\alpha}) \ln(1 - \hat{Y}_{i\ell\alpha})$$
$$\left. + \frac{1}{\sum_{i\alpha} Y_{i\ell\alpha}} \sum_{i\alpha} Y_{i\ell\alpha} (X_{i\ell\alpha} - \hat{X}_{i\ell\alpha})^2 \right] \quad (2.1)$$

with:

- l: the emission and absorption lines (16 in our case).

- $\alpha$: the "boxes", i.e the different regions of the spectrum (40 in our case)

- i: the spectra

- $\hat{Y}$ : The **predicted** classification, $\hat{Y}_{i\alpha\ell} \to 1$ if the spectrum is a LBG, 0 otherwise.

- $Y$ : The **desired** classification, $Y \to 1$ if the spectrum is a LBG, 0 otherwise.

- $\hat{X}$ : The **predicted** regression, i.e the position, between 0 and 1 of the line of the spectrum within the dedicated box.

- $X$ : The **desired** regression, i.e the position, between 0 and 1 of the line of the spectrum within the dedicated box.

The loss function combines several terms. The first two terms are similar to what is usually called a categorical cross-entropy loss. The very first term increases loss in case of false positive (LBG predicted and it's a contaminant), the second term increases loss in case of false negative (contaminant predicted and it's a LBG) and the last one is an unweighted least-squares loss increasing the loss if the predicted position in the box is different from the real one.

## 2.3 The aim

For the determination of the redshift, i.e. the regression task, the condition is to have a redshift error of ±0.05, as a very precise redshift determination phase is then carried out using a template fitting method approach. For the classification task, the objective is to obtain the best Purity/Efficiency curve, and this is where the challenge of the internship lies. An example of this plot can be seen in 5.1 for instance. Efficiency is equivalent to the classic recall and purity to the precision, but are the preferred term for the sake of physical meaning.

Efficiency is defined as:
$$\frac{TrueLBG \& LBGDetected}{LBGDetected}$$
and purity as
$$\frac{TrueLBG \& LBGDetected}{TrueLBG}$$
, with $TrueLBG$ the number of true LBGs contained in the test sample, $LBGDetected$ the number of LBGs detected by the CNN in this sample.

# Chapter 3

# The search for the best dataset with Data Augmentation

## 3.1 The initial dataset and its limitations

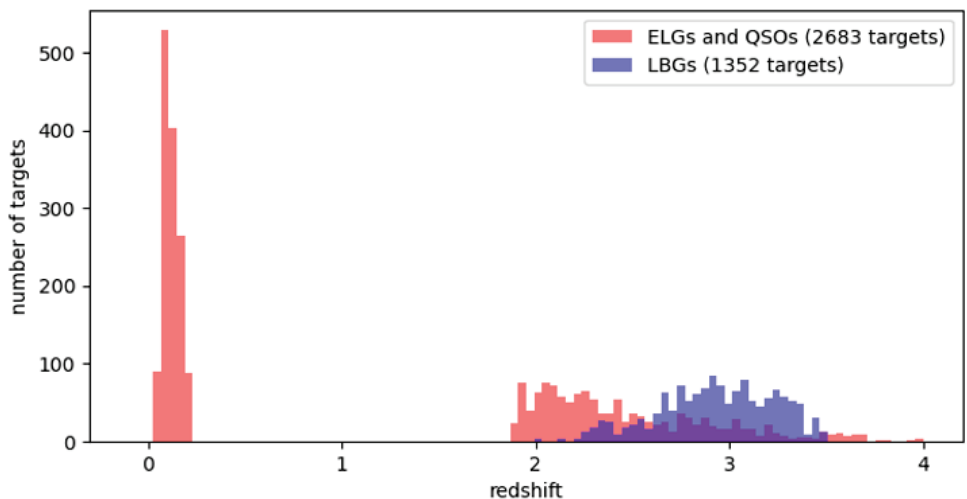

Figure 3.1: *The initial dataset.*

As with all machine learning problems, the amount of data available to train the model is undoubtedly the most important factor. The more quality data available for training, the better the training.

In the context of the problem of determining the redshift of QSOs (and not LBGs), and therefore the initial and main use of QuasarNET, the amount of data is not a problem, because the number of QSOs observed with DESI to date is very large, with hundreds of thousands of spectra available. This is not the case for LBGs, which are much less widely observed. To date, only 1.352 targets have been identified with a confidence level sufficient to ensure that they are effectively LBGs.

The initial training sample used to create the first CNN is shown in Figure 3.1. QSO and ELG spectra were introduced into the sample in proportion to the number of LBGs, so that the CNN could learn regression (redshift determination) and classification (LBG or not) simultaneously.

LBG redshifts are centered around a redshift z = 3, and range from 2.2 to 3.4.

The main limitations of this initial dataset, that can be noticed in Figure 3.1 are:

- Problem 1: Too little data, with only 1352 LBGs.

- Problem 2: Reduced redshift range, centered around 3.

By solving problem 1, we can hope to increase the efficiency of the CNN.
By solving problem 2, in particular by creating data in the redshift ranges around 2 or 4, we can hope to increase the model efficiency in these regions, which are very interesting from a scientific point of view. Indeed, if the CNN is not trained (or not trained enough) on a particular redshift region, it will be unable to deduce it on the test sample, and this has been shown by a test whose results are presented in Figure 3.2.

In Figure 3.2, we can notice that in the case where the CNN has been trained over the redshift range [2.9 - 4], and tested over this same range, we can see that the CNN has an efficiency and purity that evolves as a function of c_th, showing that the CNN has learned well to recognize LBGs over this range.

In the case where the CNN has been trained on the same amount of data, but in another range of redshifts, 2-2.9, we can see that the CNN has not learned to recognize LBGs over the test range, i.e. 2.9 - 4. Purity remains almost constant, and corresponds to the ratio of LBG contained in the test dataset, while efficiency decreases linearly from 1.0 to 0, which clearly corresponds to a random selection of LBG, regardless of the confidence threshold chosen.

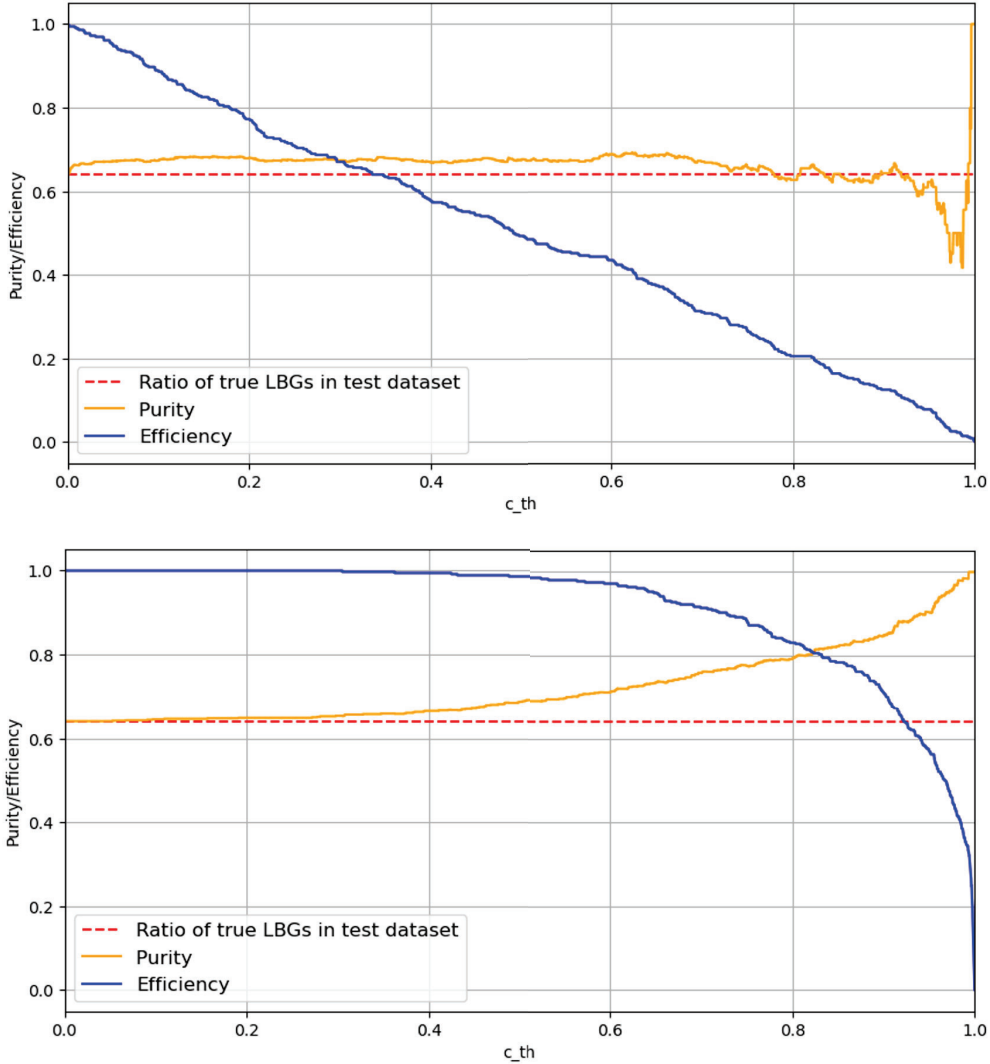

Figure 3.2: *The same QuasarNET model **trained on different datasets** and **evaluated on the same test dataset**. The redshift range of the test dataset is [2.9,4]. 1st model (Figure on the top): Redshift range of training dataset = [2,2.9], corresponding to 50% of the training dataset. 2nd model (Figure on the bottom): Redshift range of training dataset = [2.9,4], corresponding to 50% of the training dataset. The CNN returns a confidence between 0 and 1 to assess whether it is a LBG (ideal confidence of 1) or another object (ideal confidence of 0). c_th stands for "confidence threshold" and corresponds to the limit after which we consider that the CNN is labeling the spectrum as LBG. For example, a c_th of 0.8 means that we consider every spectra with a confidence level of the CNN over 0.8 as a LBG and another object otherwise.*

It was therefore necessary to find ways of increasing the size of the data sample in order to improve the model's performance, which is directly linked to it, and to increase the range of redshifts detected by the CNN.

## 3.2 Augmentation by changing the redshift

The initial dataset is centered around $z0 = 3$. We want to create a new dataset centered around z1.
We therefore want to apply a transformation to the wavelengths of the spectrum such that a redshift of $z0$ becomes a redshift of $z1$.
The relationship between redshift and wavelength is :

$$\frac{\lambda_1}{\lambda_0} = \frac{1+z1}{1+z0}$$

with $\lambda_0$ a wavelength of the regular spectra and $\lambda_1$ a a wavelength of the shifted spectra
We can rewrite as:
$$\lambda_1 = \frac{1+z1}{1+z0} * \lambda_0 \tag{3.1}$$

We then need to transform these wavelengths, which are between $\lambda_i = 3600\text{Å}$ and $\lambda_f = 10000\text{Å}$, into 800 bins between 0 and 800, e.g 1 bin $= 8\text{Å} = dl$ so that it can be given as CNN inputs.
We therefore perform the following operation:

$$X_f = (\lambda_1 - \lambda_i)/dl$$

with $X_f$ the final data, the input of the CNN.

$$X_f = ((\frac{1+z1}{1+z0} * \lambda_0) - \lambda_i)/dl$$

with equation 3.1

Figure 3.3 shows the effect of this shift towards higher and lower redshifts.

Figure 3.4 shows an example of a whole dataset shift from a mean redshift $z_0 = 3$ to a mean redshift $z_1 = 4$

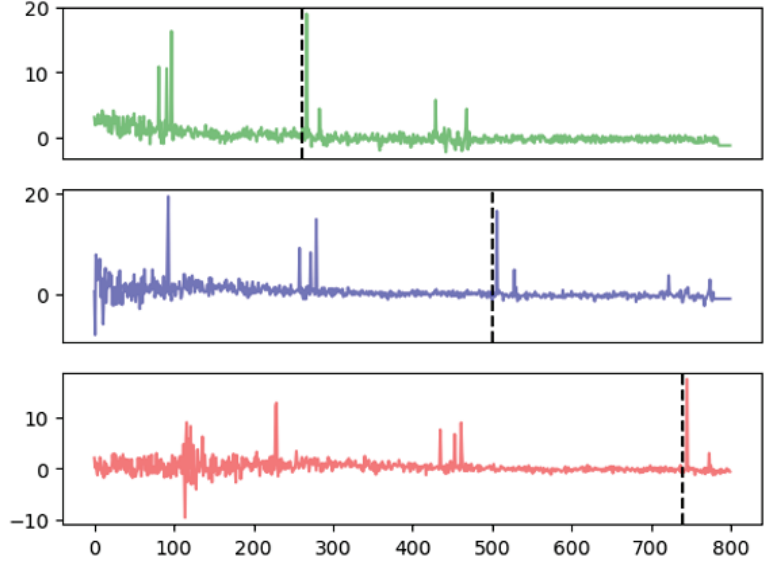

Figure 3.3: *Comparison between shifted spectrum to lower redshift with $z_1 = 2$ (green plot), and higher redshift with $z_1 = 4$ (red plot). The blue plot corresponds to the original spectrum with $z_0 = 3$.*

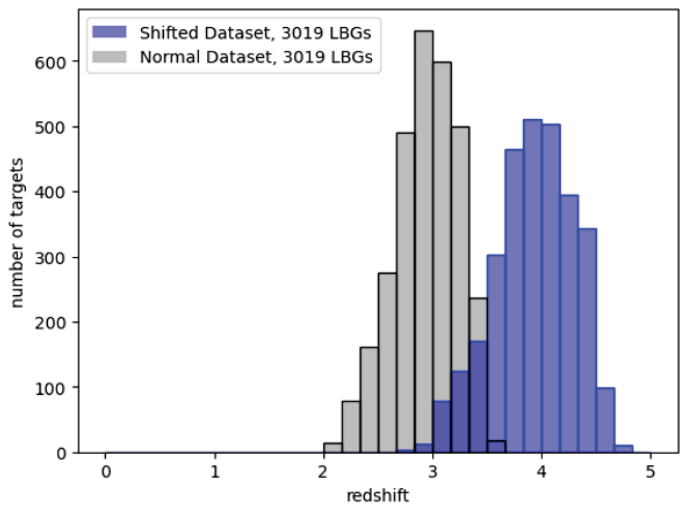

Figure 3.4: *The dataset obtained by shifting the original dataset, whose redshift is centered around $z_1 = 3$ to a dataset whose redshift is centered around $z_1 = 4$*

In order to find out whether shifting the redshift of the dataset would have any effect on the detection of LBGs in redshifts outside the range of the initial training dataset, we had no relevant test dataset. However, we were able to use a sample of "g-droupout" spectra. These LBG spectra, whose redshift is around 4, have not been classified by humans, but a fitting template method was used to determine their redshift, and thus give a redshift prediction for this sample.

The idea was therefore to compare the CNN predictions with those of RR and see if they were consistent.

The results, presented in Figure 3.5, show very good consistency between the RR predictions and those of the CNN, confirming that the CNN is now able to detect LBGs in the redshift range around 4, and predicts a suitable redshift.

However, the area surrounded by the red circle is rather intriguing. It would appear that between 3.75 and 3.8, the CNN is wrong on the predicted redshift, which poses a problem. In fact, we realized that this was linked to a deeper problem, which led to the idea of masking an area of the spectrum, explained in the next section.

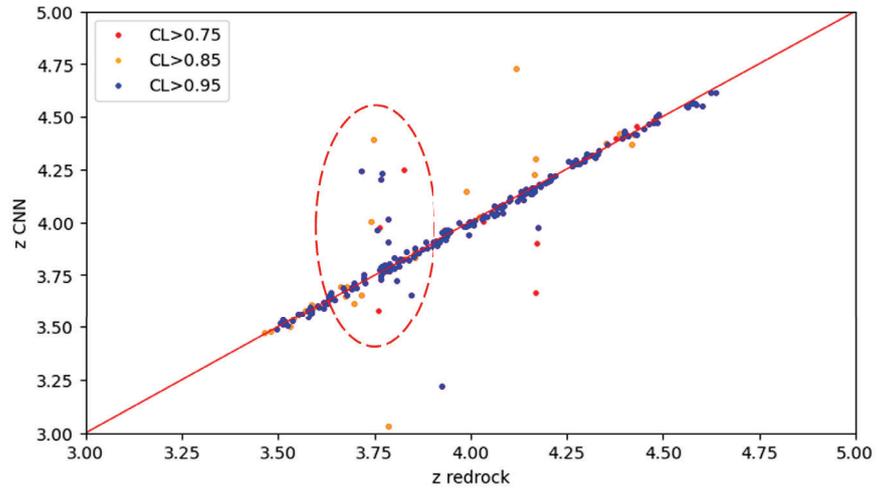

Figure 3.5: *Comparison between the redshift predictions of Redrock and the CNN trained with shifted data for a sample of g-dropout LBGs whose redshifts is centered around 4. CL stands for "Confidence Level" and corresponds to the minimum confidence level required to consider the assumption of the CNN as True. The higher CL, the tighter the selection of LBGs by the CNN.*

## 3.3 Augmentation by masking some parts of the spectra

### 3.3.1 Masking a critical region of the spectrum

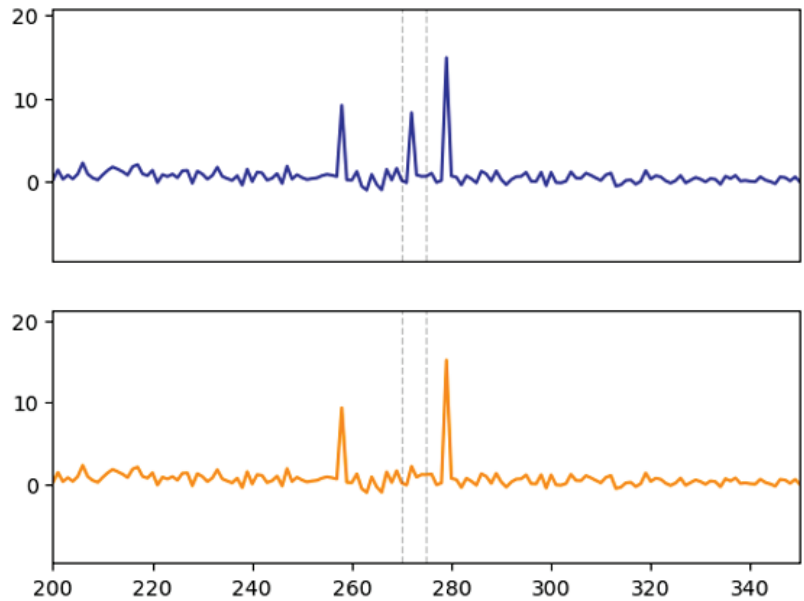

Figure 3.6: *On the top (dark blue plot), the original spectrum. On the bottom (orange plot), the same spectrum with a 40Å range (between the vertical dashed lines) replaced with noise.*

The spectrum obtained is the superposition of the flux from 3 different spectrographs, as shown in Figure 3.7. There are zones of superposition, and in particular the 5800 Å(i.e 580 nm on the figure) zone of overlap between the two spectrographs. In this zone, the flux of both spectrographs is very low, and the signal is consequently very noisy.

The concern mentioned at the end of Section 3.2 is related to this.

Indeed, when we look at the redshift corresponding to the superposition of this zone with the Ly-$\alpha$ line, the most characteristic line in the LBG spectrum, we find :

$$z_{pb} = \frac{\lambda_{pb}}{Ly\alpha_{rf}} - 1$$

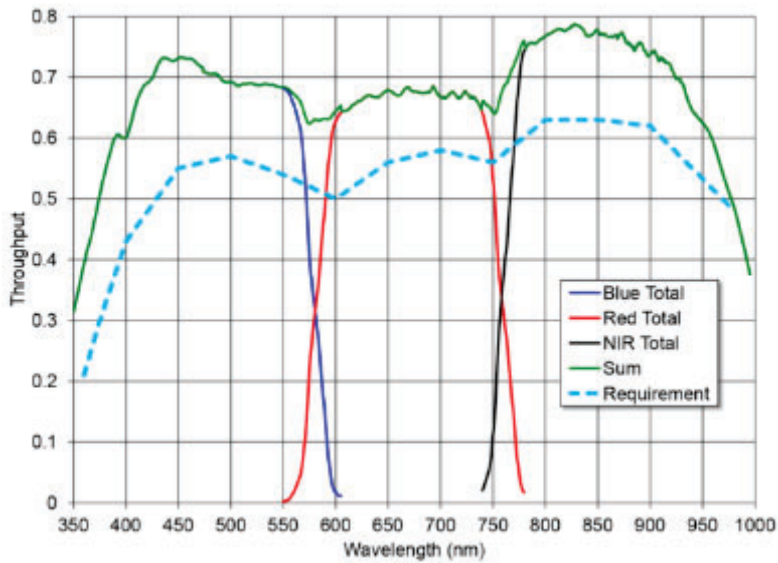

Figure 3.7: *The flux distribution of the 3 spectrographs of DESI with respect to wavelength. From [8].*

With $\lambda_{pb} = 5800 \text{Å}$, the wavelenght of the overlap problem, and $Ly\alpha_{rf} = 1215.67 \text{Å}$, the wavelenght of the Ly$\alpha$ line in its rest frame, e.g at $z = 1$.

$$z_{pb} = 3.77$$

We then find that the redshift corresponds well to the redshift range of the problem presented in Figure 3.5, which is $[3.75 - 3.80]$.

Due to the black-box nature of the neural network, it's difficult to know exactly what's behind this. One explanation could be that the CNN relies heavily on the Ly$\alpha$ line to determine redshift, looking for it in the very noisy zone when it comes to redshift around $z = 3.77$, and confusing it with a false line in the same region, created by noise.

To solve this problem, the idea was to mask this area of overlap between the two spectra, by adding Gaussian noise, as shown in Figure 3.6, where the zone between the two dotted lines was masked. After several trials, the optimum width of the zone was determined to be 40 Å, and gave satisfactory results, shown in Figure 3.8. We can see that this time, redshift values around 3.77 are much closer to the red line than before, the outliers have been eliminated overall, without deteriorating the spectrum too much.

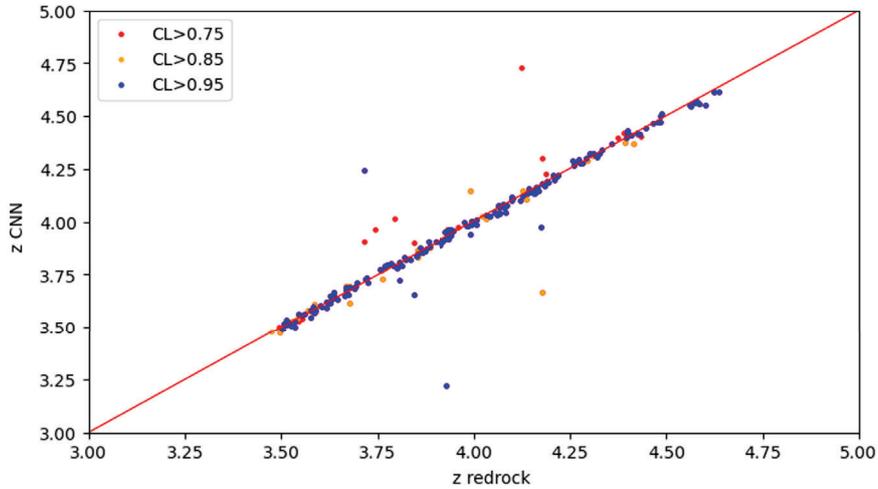

Figure 3.8: *Comparison between the redshift predictions of Redrock and the CNN trained with shifted data for a sample of g-dropout LBGs whose redshifts is centered around 4. Compared to Figure 3.5, the CNN was trained with a dataset whose region of 40 Å around 5800Å has been masked.*

### 3.3.2 Masking the Lyman-$\alpha$ emission line

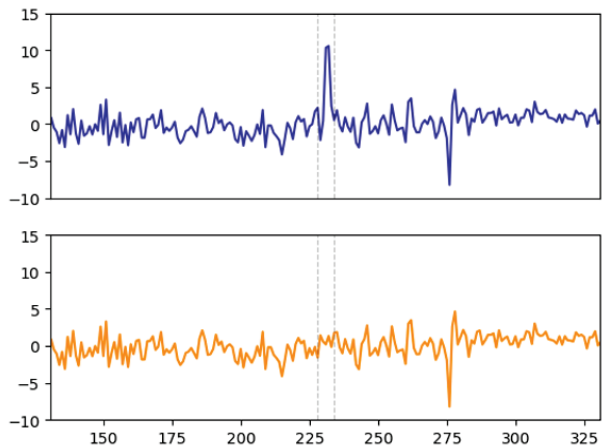

Figure 3.9: *On the top (dark blue plot), the original spectrum. On the bottom (orange plot), the same spectrum with a 40Å range (between the vertical dashed lines) replaced with noise, corresponding to the position of the Lyman-$\alpha$ emission line.*

On the same principle, we came up with the idea of masking the Ly$\alpha$ line to create a new dataset that is again different from the initial one, and to force the CNN to learn the LBG spectrum without the Ly$\alpha$ line to which it seems to attach so much importance.

This is important because, while in the case of LBGs that are LAEs, the Ly-$\alpha$ line is sufficient to determine the redshift of the LBG, this is not the case for LBGs with a small or no Ly-$\alpha$ line emission. In Figure 3.9, the Ly-$\alpha$ line is clearly present in the upper spectrum, while it disappears in the lower spectrum.

## 3.4  Augmentation by adding noise

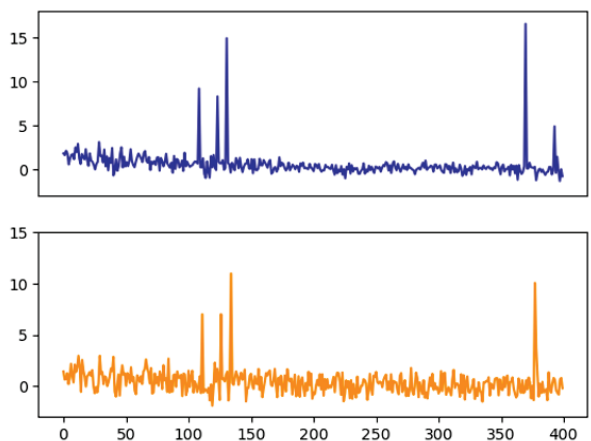

Figure 3.10: *On the top (dark blue plot), the original spectrum. On the bottom (orange plot), the same spectrum with Gaussian noise added across the spectrum.*

In a similar way to adding noise or blur to an image for data augmentation, another idea that came naturally was to add noise to the spectra. This was combined with a slight redshift to avoid recreating spectra too close to the original. Figure 3.10 shows the addition of noise to the entire spectrum.

These spectra were used in the final best dataset presented in Section 3.6.

## 3.5 Adding synthetic spectra

### 3.5.1 The four templates

To increase even more the size of the dataset, we have been able to use artificially created spectra mimicking real spectra obtained after 120min exposure. A total of 25,200 spectra could be used, enabling a significant increase in dataset size. These spectra are divided into 4 different templates, shown in Figure 3.11. The purpose of these 4 templates is to represent the different types of LBG spectra in existence, with more or less pronounced Ly-$\alpha$ emission lines. We can see that for Template 0, the Ly-$\alpha$ line is completely absent, whereas on Template 3, it is very pronounced. The aim is to reproduce the different types of LBG spectra that can be encountered.

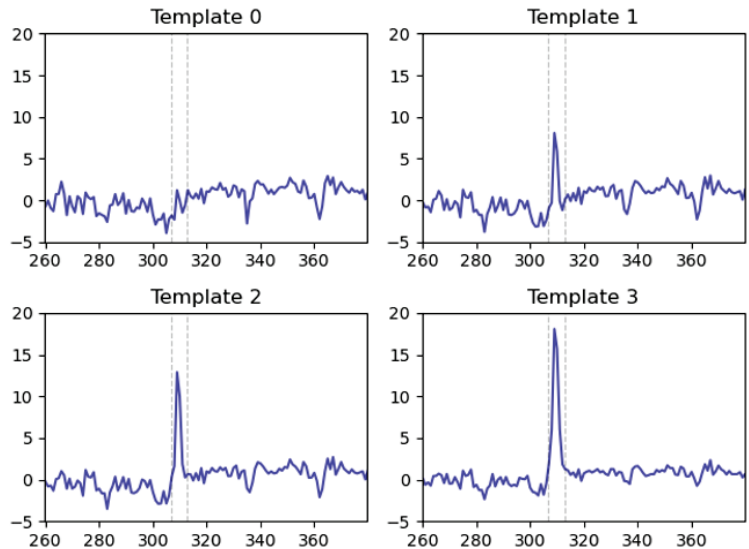

Figure 3.11: *The four templates of synthetic spectra. The higher the template number, the more pronounced the Ly-$\alpha$ line.*

Benefits of synthetic spectra:

- The redshifts are ranging from 2 to 4, evenly distributed. The large number of redshifts also enables to create a reliable test dataset with previously difficult-to-reach redshift ranges. It was thus possible to compare the different templates on an independent test dataset, particularly at interesting redshifts around 2 and 4.

- 4 templates and 6 different magnitude levels, increasing the diversity in the spectra.

- Spectra very close to reality, with noise created using noise patterns from DESI sky noise.

Disadvantages of synthetic spectra:

- There is a slight shift in the simulation of the Ly-$\alpha$ line. In real spectra, a part of the Ly-$\alpha$ line is not visible, and this effect has not been considered in the simulations.

- Although it mimicks pretty well real spectra, it turned out to differ a bit. Indeed, for a big part of the internship, the models were compared on a test sample taken from the simulations, but we realized that the results were not exactly the same as on a test sample taken from the real data.

- The spectra are binned in redshift every 0.1, and this poses a problem presented in Section 3.5.2.

### 3.5.2 Getting a continuous redshift distribution

The simuated spectra are binned in redshift and their redshift distribution is therefore not continuous. This poses a major problem because, as shown in Section 3.1, the CNN is only effective on redshift ranges it has already seen during training. In this case, the redshift range is limited to the 20 discrete different redshifts distributed between 2 and 4. We can expect very good efficiency at these exact redshifts (or at redshifts very close), but not between them.

The solution that was then implemented to face this problem was to create from this dataset with discrete redshifts a dataset with continuous redshifts, using the methodology presented in Section 3.2

Figure 3.12 shows the result of this transformation. To add a realistic effect to the redshift distribution, a random side was added to the redshift change. We can see that in the final dataset (orange distribution), the redshift distribution is smooth and thus more convenient for the CNN training.

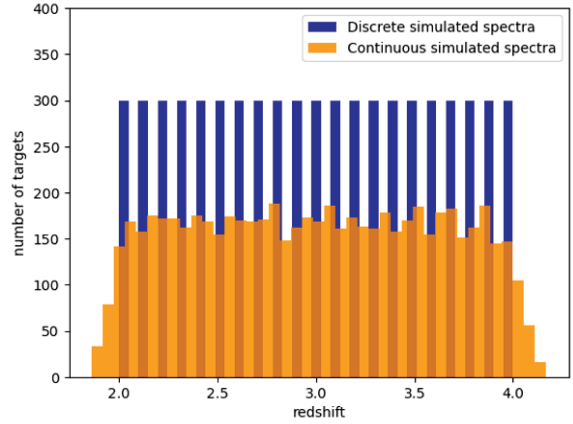

Figure 3.12: *The redshift distribution of the initial synthetic spectra (discrete distribution) and of the final distribution (continuous distribution).*

## 3.6 Selection of the best dataset

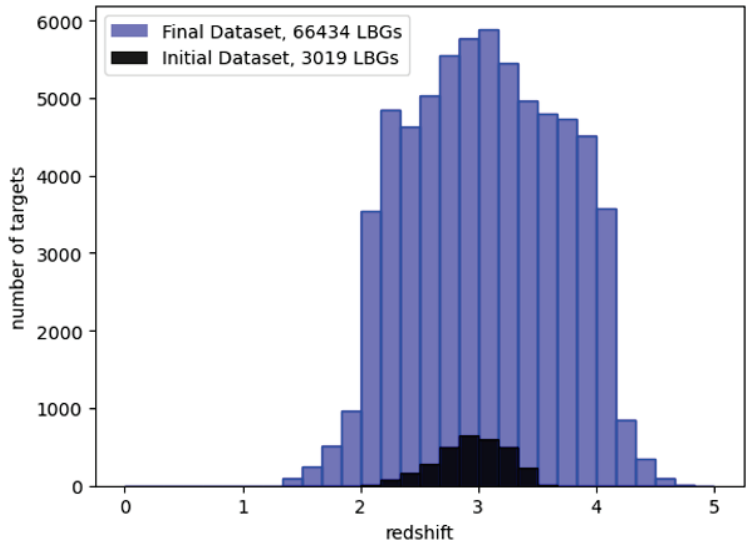

Figure 3.13: *Comparison between the redshift distribution of the initial dataset (black histogram) which contains 3019 LBGs and the final, augmented one (blue histogram), which contains 66434 LBGs*

In Figure 3.13, we can see that with all these manipulations, the number of spectra available for CNN training has risen from 3019 to 66434, an increase of around 22x, as far as LBGs spectra are concerned. The range of redshifts is also wider and less disparate than initially.

Given that the data increase was very substantial, and even if the modifications are multiple, the modified spectra are taken from the same initial dataset. These 66434 spectra are taken from 20 different samples, which may or may not have undergone different noise, redshift or other modifications.

The idea was then to create the most effective training dataset for the CNN regression and classification tasks from these 20 samples. To this end, 16 different training datasets were created from different combinasions from these 20 samples and compared on the same independent test dataset. The results are available in Appendix C.4, C.5, and C.6.

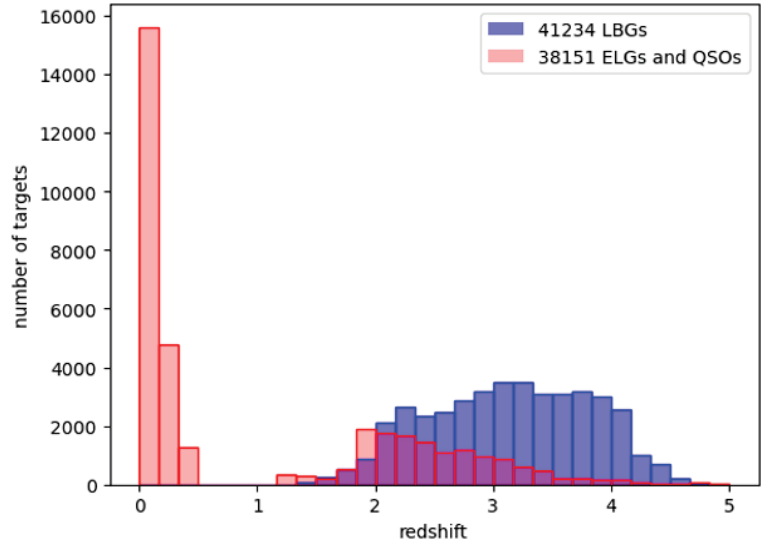

Figure 3.14: *The redshift distribution of the best dataset for a QuasarNET-like CNN model with regards to an independant test dataset.*

Figure 3.14 shows the dataset that stood out as the most efficient one. We can see that for the LBGs, most of the targets have redshifts in [2,4]. To create this dataset, the following 8 samples are used:

- The initial dataset with noise in the overlap area

- A dataset whose mean redshift has been shifted from $z0 = 3$ to $z1 = 4$

- The same dataset, keeping only redshifts above 2.7

- A dataset whose mean redshift has been shifted from $z0 = 3$ to $z1 = 2$

- A dataset with artificially added noise slightly shifted from z0=3 to z1=3.3

- All simulations from the 4 different templates

- QSOs with redshift above 2

- ELGs with redshift in [0.05,0.15].

Regarding the choice of QSOs and ELGs, several LBGs/(QSOs+ELGs) ratios were tested. If there are too many QSOs and ELGs compared with LBGs, the CNN may be better suited to its Classification task than to its Regression task, and vice versa. 48% (38151 ELGs + QSOs/79385 Targets) is a good trade-off between these two tasks regarding different tests that have been conducted to this end.

The range of redshifts for these two types of objects was not chosen at random either. As far as its classification task is concerned, the CNN will ultimately have to distinguish LBGs from specific QSOs and ELGs. Indeed, to determine which objects to observe, a so-called "target selection" stage aims to use a photometric survey to identify LBGs. However, QSOs with a redshift greater than 2, and ELGs with a redshift in [0.05, 0.15] are indistinguishable from LBGs in these photometric surveys, hence the classification role of the CNN is to separate these specific QSOs and ELGs from LBGs using the observed spectra.

# Chapter 4

# Model's architecture tuning

## 4.1 Transfer Learning

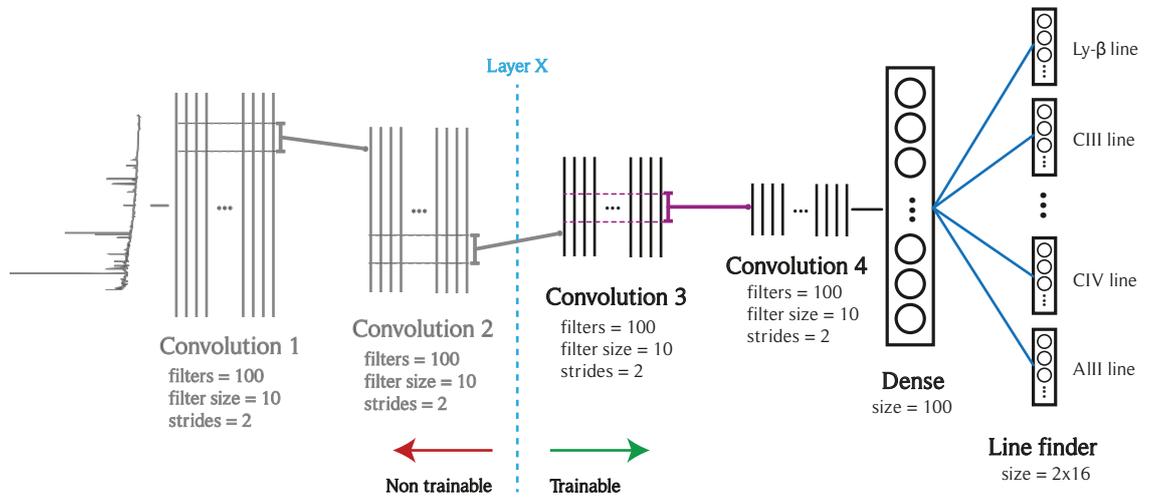

Figure 4.1: *The principle of Transfer Learning on a QuasarNET architecture, with the X first layers set to non-trainable. Only the most important layers have been represented.*

Although the dataset size is now much larger than initially, it is still much smaller than that used in [7] to determine the redshift of QSOs. For this application, the QuasarNET model has been trained on hundreds of thousands of QSO spectra, and produces overall better results than the QuasarNET model adapted for LBGs.

In this kind of case, a widely-used method, particularly in image processing, is the so-called Transfer Learning (TL). Transfer Learning is the technique of using the knowledge of a pre-trained model on one task to improve performance on another, similar task. For example, a model trained on many truck images could be reused to recognize cars, by re-training only part of the model.

The idea was therefore to apply this principle to the QuasarNET weights resulting from training on QSOs.
The principle is explained in Figure 4.1. First, we initialize the weights trained on quasars in a QuasarNET model.

Then, we set these weights to "non-trainable" mode up to a certain layer, and "trainable" mode for the weights located downstream of this layer, which, unlike the first ones, will be trained. Once this training is complete, it's also possible to run a few epochs leaving the entire model in trainable mode.

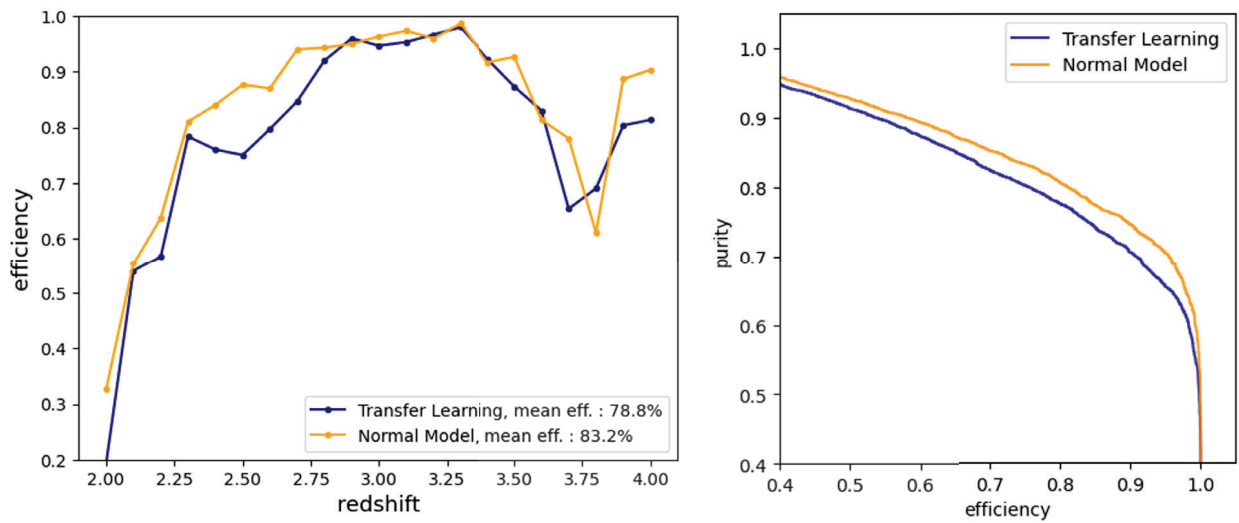

Figure 4.2: *Comparison between the best Transfer Learning model (dark blue) and the normal model (orange) trained on the same training dataset.*
*Right hand side figure: Purity/Efficiency plot for a test dataset of synthetic spectra.*
*Left hand side figure: Efficiency of the models regarding the redshift of the test dataset.*
*"mean eff." stands for the mean efficiency over the whole range of redshifts.*

In this way, it is hoped that the first layers that have been put into "non-trainable" mode will be more efficient than before, as they have been trained on a much larger amount of data, but that the model will remain adapted to the LBG problem, as the downstream "trainable" layers have been trained on LBG spectra. Numerous tests were carried out, varying the number of trainable and non-trainable layers.

Figure 4.2 shows the results of the best model, with 7 layers set to non-trainable for the best Transfer Learning model. The graph on the right shows that the TL model is generally a bit worse than the normal model in terms of efficiency/purity ratio, since the blue curve is below the orange curve. In the graph on the left, we can't see any noticeable improvement over an interesting redshift range, and overall the TL model results are a bit worse than the normal model for this test dataset.

As a conclusion, the best TL model turned out to be slightly worse than the normal model, even though they were trained with the same dataset (from the data augmentation presented in Chapter 3). However, it would be interesting to imagine some TL on a larger model, notably with more layers, such as the one presented in Section 4.4. These results are encouraging, since quasar spectra are very different from those of LBGs, and the results obtained are acceptable compared with those of the classical model.

## 4.2 End-to-end learning

End-to-end learning is a method that gives the deep learning model more freedom, by integrating less knowledge about the features (such as the link between line position and redshift in this case) and letting the model directly determine an output based on the input by managing all the steps, hence the name end-to-end.

In the case of QuasarNET, the model is trained to identify the various lines present in the spectrum of a LBG. We then takes the position of the line for which the model is most confident, and deduces the redshift.

While this works perfectly when the model is very confident on the best line, in more complex cases this solution might not be optimal. This is because we retain only the information from the prediction on the best line, whereas the CNN has predicted the position of all the other lines, with higher or lower confidence thresholds.
The idea is therefore not to consider only the best line detected by the CNN, but to consider all the lines at once.

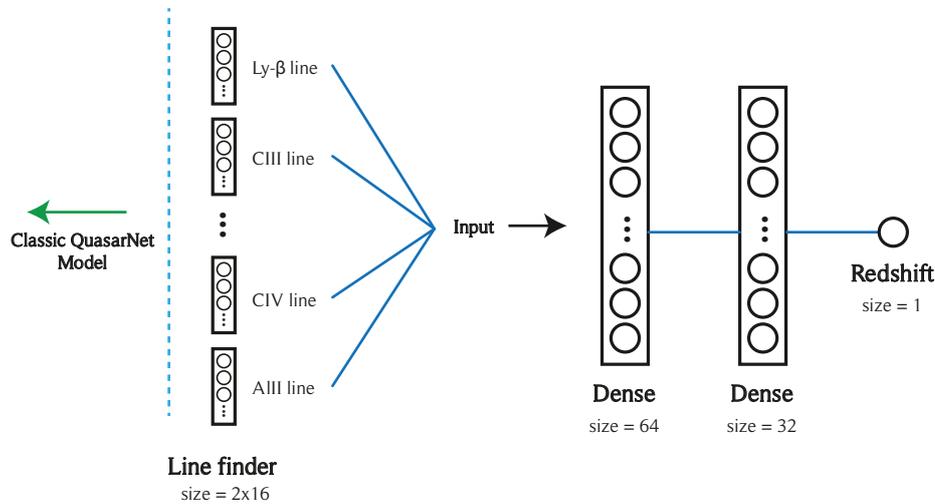

Figure 4.3: *First idea: the architecture of the classic model whose output is given as input to a Fully Connected Neural Network.*

To achieve this, two ideas were tested:

1. **An additionnal Neural Network** at the end of the classic model.
   This idea is shown in Figure 4.3. It involves adding a small Neural Network to the output of QuasarNET's Line Finder to automatically determine the information of interest, the redshift, from the positions of the lines and their confidence level. Despite several trials, the model was clearly less efficient than those of the conventional method, and are presented in Appendix D.3.

2. Doing away with the line finder and **directly deducing redshift**
   The term "end-to-end learning" is more appropriate here, since the idea is to do away with the line finder and replace it with a fully connected neural network for direct redshift inference. A representation of the model architecture is shown in Figure 4.4 The input is therefore the spectrum and the output the redshift, without adding any knowledge of the position of the lines. Again, the results were worse than with the classical model, but still encouraging. These results are presented in Appendix **??**.

These two ideas were tested only on the CNN regression task, but could have been adapted to incorporate the classification task as well. It would therefore be interesting to try out other architectures with this same principle, if we are to have any hope of improving accuracy.

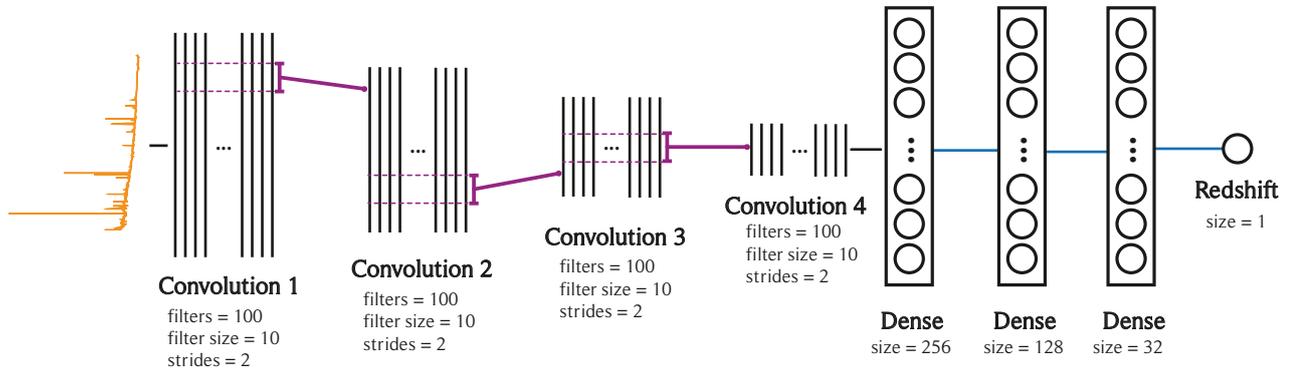

Figure 4.4: *Second idea: the architecture of an end-to-end model, directly giving the redshift as output.*

## 4.3 The Signal-to-noise ratio as additional input

The spectra given as input to the CNN are derived from the flux of the 3 different spectrographs, weighted by the noise. However, the flux and noise information are concatenated into a single dataset, so there is a loss of information.

To remedy this, we created an alternative version of QuasarNET, shown in Figure 4.5. Analogous to an image with different channels: Red, Green, and Blue, here the model takes as input not only the spectrum, but also the Signal-to-Noise Ratio (SNR). The

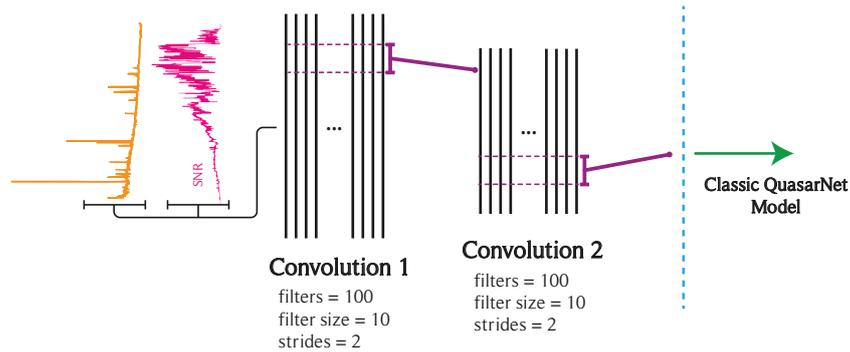

Figure 4.5: *The architecture of a classic QuasarNET model, with the SNR as additional input.*

SNR is defined as
$$SNR(\lambda) = \sum_{spec.}(flux * \sqrt{ivar(\lambda)})$$

with $\lambda$ the wavelenght, spec. the different spectrographs and ivar is the inverse of the variance of the flux, i.e the noise.

The SNR is used to measure the extent to which the data is relevant or not. The higher the SNR, the higher the quality of the data.

## 4.4 Hyper-parameters tuning with Bayesian Optimization

### 4.4.1 Classic model's hyperparameters search

A hyperparameter is an external parameter that is defined before the start of the model learning process. Unlike the model parameters themselves, which are learned from the data during the learning process, hyper-parameters are variables that determine how the learning process is to be executed. They influence model performance but are not learned automatically, which makes the difference between parameters and hyper-parameters. The learning rate, the number of layers or the filters size are, in our case, example of hyperparameters.

The aim is therefore to determine the best hyper-parameters. A first solution is to test different values and hope to see improvements, but this technique can be slow and very costly in computing ressources. To overcome this problem, optimized algorithms for finding hyper-parameters have been developed (See [5] for more details). The one we used in this case is called Bayesian Optimization.

Bayesian optimization is an optimization method that uses probabilistic models to find the best hyper-parameters. It exploits an iterative approach based on probabilistic models to efficiently search for optimal hyperparameters by sequentially evaluating different configurations with balanced exploration-exploitation (See [5]) .

The hypermodel will perform a few epochs with a randomly selected set of hyperparameters, within the given data range, to see whether or not learning seems to be efficient. Regarding the result of this training, it will then change its hyper-parameters and reiterate, until it finds the hyper-parameters that give the best learning.

| Hyperparam. | Search range | Before search | 1st search | 2nd search |
|---|---|---|---|---|
| nbr layers | [2,20] (int) | 4 | 8 | 10 |
| nbr filters | [10,500] (int) | 100 | 323 | 420 |
| learning rate | [0.0001,0.1] (float) | 0.001 | 0.01 | 0.1 |
| strides | [1,8] (int) | 2 | 2 | X |
| activation fct | [sigm., r. sigm., lin.] | rescaled sigmoid | sigmoid | X |
| dropout rate | [0,1] (float) | 0 | 0.19 | X |

Table 4.1: *2 hyperparameters search. The 1st search was conducted on 90 trials and the second one on 25. The X represents parameters that have not been searched. "sigm." stands for sigmoïd, "r. sigm." for rescaled sigmoïd, "lin." for linear, "fct" for function and "nbr" for number.*

Two Bayesian optimizations for hyper-parameter searches have been carried out on the QuasarNET base model. The results of these two hyperparameter searches are presented in Table 4.1. Overall, the results are in favor of a larger number of layers and filters. This considerably increases the training time, and results in a much larger model, with more parameters to train.

A new architecture has been created from this result, presented in Figure 4.6. We can see that compared to the basic model (Figure 2.2 to see the initial architecture), one of the biggest changes are the 10 convolution layers with a filter size of 420.

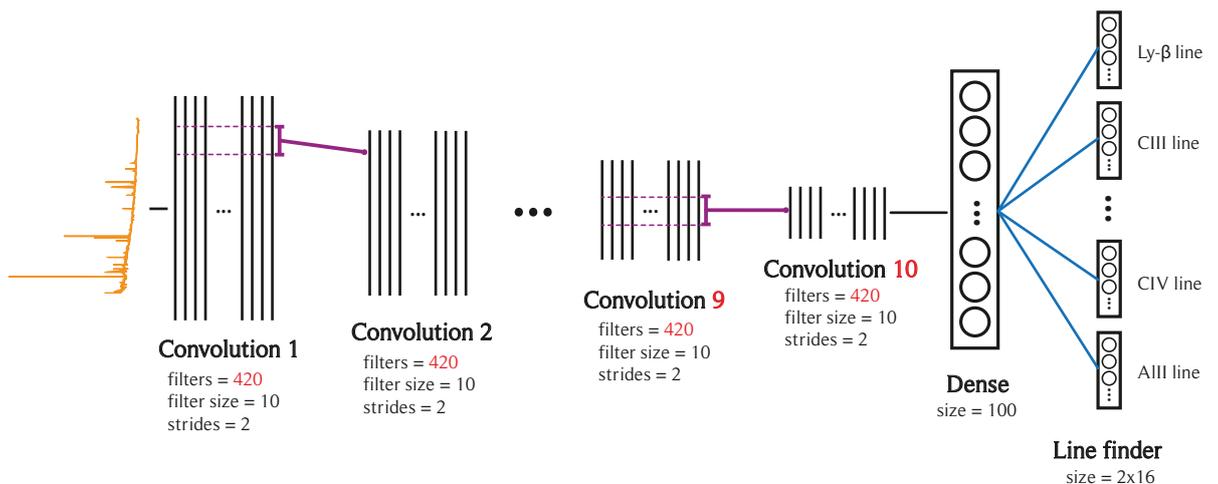

Figure 4.6: *The architecture of a QuasarNET model whose hyperparameters have been set to the best ones found by the Bayesian Optimization.*

### 4.4.2 SNR model's hyperparameters search

Following exactly the same principle, a search for hyperparameters was carried out for the SNR model. The results of this search are shown in Table 4.2. Once again, the size of the model and the number of parameters increase, slowing down the learning process.

| Hyperparam. | Search range | Before search | 1st search | 2nd search |
|---|---|---|---|---|
| nbr layers | [2,20] (int) | 4 | 4 | 8 |
| nbr filters | [10,500] (int) | 100 | 296 | 173 |
| learning rate | [0.0001,0.1] (float) | 0.001 | 0.001 | 0.001 |
| activation fct | [sigm., r. sigm., lin.] | rescaled sigmoid | linear | X |
| dropout rate | [0,1] (float) | 0 | 0.05 | X |

Table 4.2: *2 hyperparameters search for the "SNR" model. The 1st search was conducted on 22 trials and the second one on 9. The X represents parameters that have not been searched. "sigm." stands for sigmoïd, "r. sigm." for rescaled sigmoïd, "lin." for linear, "fct" for function and "nbr" for number.*

For both kind of models, the larger the list of hyper-parameters to be tuned, the larger the search space, and the slower the search. It is therefore advisable to keep the number of parameters to a minimum when searching for hyper-parameters. Not all parameters could be tested, but it would also have been interesting to carry out this search with other hyper-parameters such as the number of boxes (nb_boxes), which corresponds to the number of zones in which the CNN will search for lines in its Line Finder, or the coefficient in the loss function (see Section 6.3).

# Chapter 5

# Final models and results

## 5.1 Final results on Purity and Efficiency

This section presents the results of the best-performing models, on 3 different test datasets. A score has been assigned for each test, corresponding to the point where Purity = Efficiency.

The four models tested were

- **Initial:** The initial model at the beggining of the internship, without data augmentation or architecture changes.

- **Classic:** The initial architecture, but with optimal data augmentation (see Section 3.6).

- **Hyp Classic:** The architecture resulting from hyperparameters tuning (see Section 4.4.1) with optimal data augmentation.

- **Hyp SNR:** The architecture including an SNR layer derived from hyperparameters tuning (see Section 4.4.2) with optimal data augmentation.

3 tests were carried out to evaluate the performance of these 4 models.

The first test was carried out on a dataset that was set aside from the training of all the models. It is made up of LBGs with a redshift centered around z=3, whose distribution is presented in Appendix C.1, just like the training data. All augmented versions of these spectra (shifts, noise addition, etc.) have also been removed from the training data. The results of the different models on this test dataset are shown in Figure 5.1.

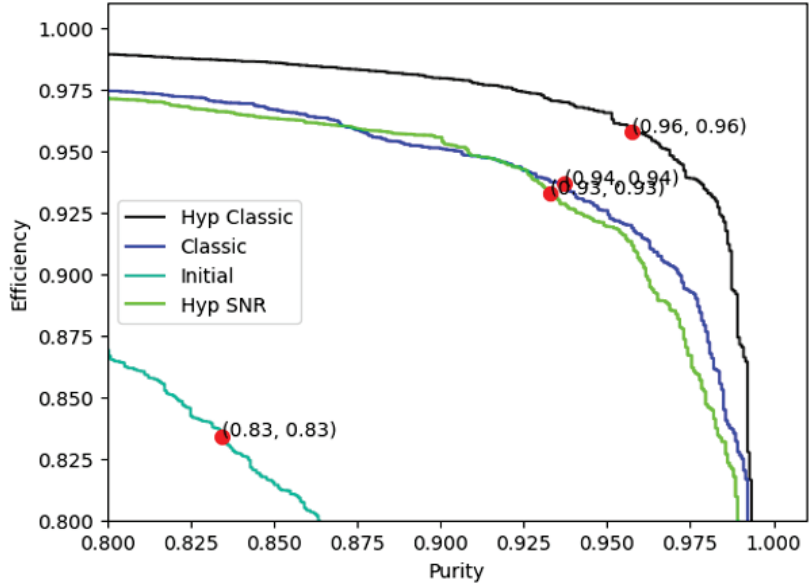

Figure 5.1: *Purity/Efficiency plots of the initial and the 3 best final models, on a test dataset of 1667 real spectra (different from training dataset) in redshifts centered around 3 ( 95% of $z \in [2.6, 3.4]$ )*

It can be seen that the model that stands out above the others is Hyp Classic, whose curve is above the other models. Compared with the initial model, we go from a score of 0.83 to 0.96, a gain of 13%. The Hyp SNR model, which is overall the most interesting model, achieves a gain of 10

The other two tests were performed on new synthetic spectra, which were not used during training.

On the one hand, a test was carried out on low redshifts, i.e. $z <= 2.6$, the results of which are shown in Figure 5.2.

It can be seen that the Hyp SNR model curve is slightly above that of the Classic model, and well above the other two. In terms of score, it goes from a score of 0.72 for the Initial model to a maximum score of 0.94 for Hyp SNR, a gain of 22%.

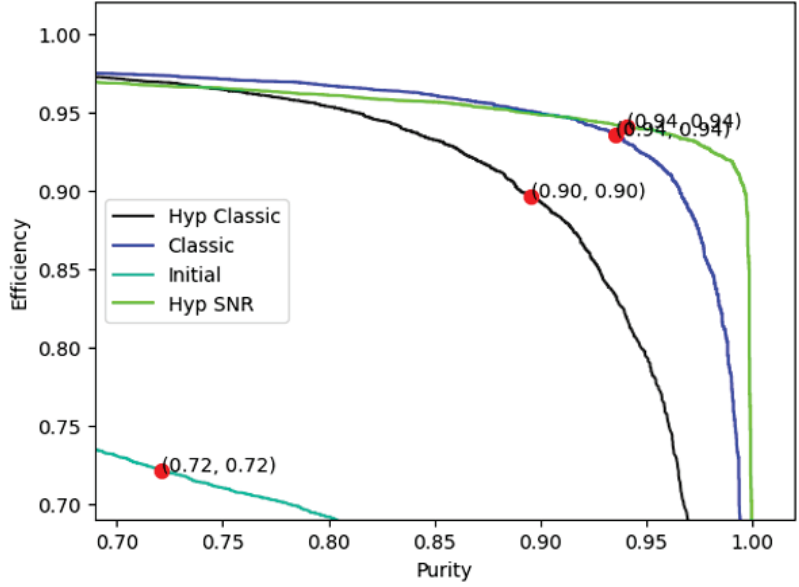

Figure 5.2: *Purity/Efficiency plots of the initial, and the 3 best final models, on a test dataset of 6622 synthetic spectra in low redshifts ($z <= 2.6$).*

A test was also carried out on high redshifts, i.e. $z >= 3.4$, the results of which are shown in Figure 5.3. The figure on the right is an enlargement of the figure on the left.

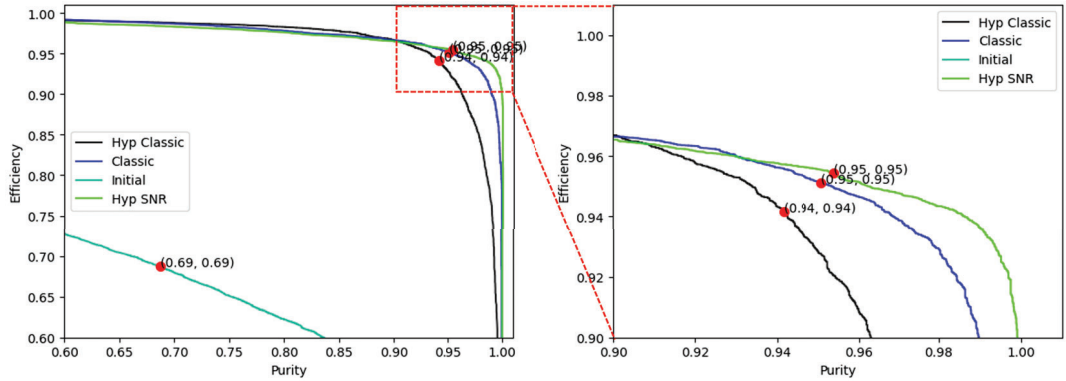

Figure 5.3: *Purity/Efficiency plots of the initial and the 3 best final models, on a test dataset of 8956 synthetic spectra in high redshifts $z >= 3.4$.*

Once again, we can see that the Hyp SNR model curve is above those of the Classic model, which comes in second. This time, the score rises from 0.69 for the Initial model to 0.95 for Hyp SNR, a gain of 26%.

The smallest gain is in the redshift range around 3, which is expected as the Initial model was trained on data with the same distribution.

Overall, the Hyp SNR model is undoubtedly the most interesting, as it achieves maximum gain in the low and high redshifts, which are the areas of greatest scientific interest.

## 5.2 Closer look on model architectures

| model | nb_layers | nb_filters | lr | nb_params | score |
|:---:|:---:|:---:|:---:|:---:|:---:|
| Initial | 4 | 100 | 0.01 | $788 * 10^3$ | [0.72,0.83,0.69] |
| Classic | 4 | 100 | 0.01 | $788 * 10^3$ | [0.94,0.94,0.95] |
| Hyp Classic | 10 | 420 | 0.1 | $16.6 * 10^6$ | [0.90,0.96,0.94] |
| Hyp SNR | 8 | 173 | 0.001 | $2.45 * 10^6$ | [0.94,0.93,0.95] |

Table 5.1: *Comparison between the architecture parameters of the initial model and the 3 best-performing models, and their scores on the 3 tests presented in Section 5.1*

The table 5.1 shows a comparison between the four models architecture parameters. nb_layers and nb_filters correspond respectively to the number of convolution layers in the model, and to the number of associated filters. Lr corresponds to the learning rate, and nb_params to the total number of model parameters, which is roughly equal to the number of trainable model parameters. This parameter is important because the greater is the number of parameters, the longer it will take to train and predict the model.

Finally, the score corresponds to the notation presented in SECTION X, on the first, second and third tests.

It can be seen that the Hyp Classic model stands out by far with the highest number of hyperparameters, without standing out as the best model overall, with notably a third position on the two tests at high and low redshift.

The Hyp SNR model seems to offer the best compromise. It has a greater number of parameters than the initial architecture, notably due to its higher number of layers, but by a factor of 3.1, well below that of Hyp Classic, which is 21. It offers the best results on both high and low redshift tests, and is competitive on classic redshifts around 3. It would therefore seem natural to opt for this choice in future predictions for DESI.

The Classic model has fewer parameters than Hyp SNR and Hyp Classic, but still delivers competitive results, and may also be interesting to use in cases where training or prediction must be carried out with limited computing power or time resources.

# Chapter 6

# Areas for improvement and curiosities

Unexpected results were observed during this project, but solving the potential problems associated with them was beyond the scope of this internship. These are thus avenues for improvement, as understanding them will possibly lead to improvements in the model.

## 6.1 Uneven importance of lines detection

For its regression task, the CNN assigns each emission or absorption line in the spectrum a position and a confidence between 0 and 1 associated with that position. The more confident the CNN is to locate a line in a particular region of the spectrum, the closer the value will be to 1, and vice versa. We then choose the line for which the CNN returns an output closest to 1, i.e. the line for which it is most confident. It is from this "best" line that the redshift of the spectrum is established.

If we look at a sample of tests and see how often the CNN finds the various lines most confidently, we would expect to find the most recognizable Ly-$\alpha$ emission line in first position, and an overall equitable distribution between the other lines.

However, when we perform the test, the results of which are shown in Figure 6.1, we realize that this is not the case. It would appear that the CNN gives decreasing importance to the different lines, and does not seem to give particular importance to the Ly-$\alpha$ emission line.

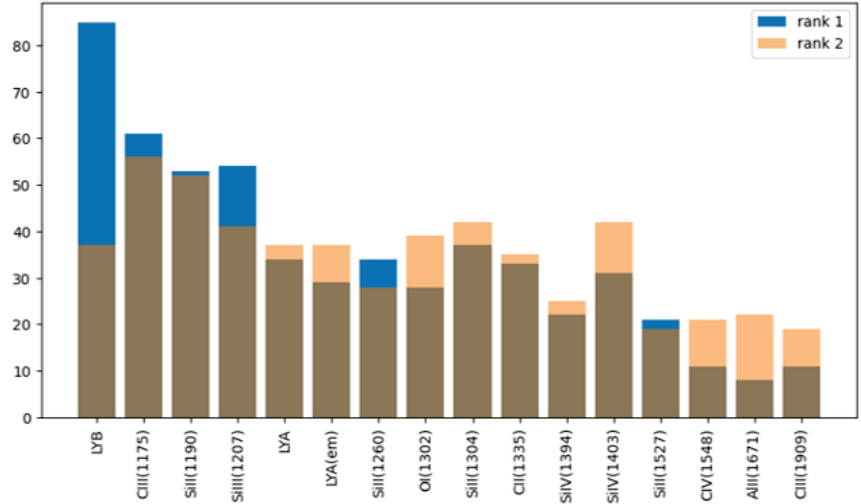

Figure 6.1: *For each line, the number of times it has obtained the maximum confidence threshold (blue histogram) or the second maximum (orange histogram) over a test dataset. The higher the histogram for a line, the more the CNN relies on the position of this line to determine the redshift of the spectrum.*

Moreover, there is another similar result regarding QuasarNET's loss function. In this loss function, the errors for each line position in the spectrum are added together. The aim is then to reduce this loss function during training, by adapting model weights accordingly.

However, when we look at a training session, presented in Figure 6.2, it seems that the various lines losses are not equivalent. We can see that the loss function value associated with the first line, named "conc_box_0", is higher than that associated with the second, conc_box_1, and so on down to the last line.

To sum up, it seems that the model gives more importance to the first lines than to the last ones, at least for the classic training dataset, and it would be interesting to understand why.

## 6.2 Learning a noise pattern

A second test was performed thanks to the variety of synthetic spectra. In particular, these spectra have different magnitudes, and it is expected that the greater the magnitude, the less the noise in the spectrum, the better the prediction quality of the

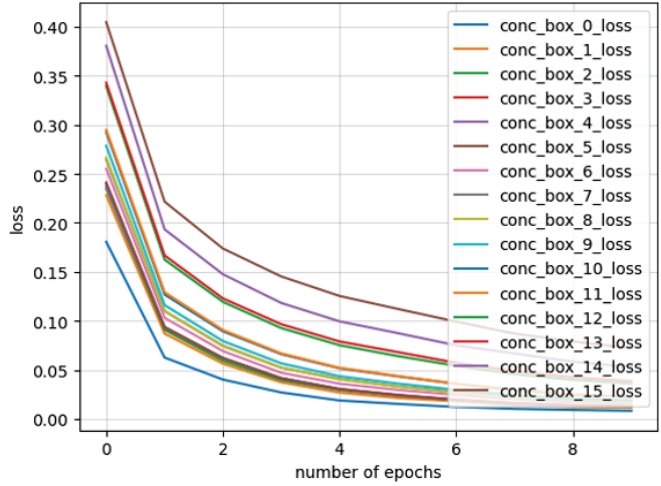

Figure 6.2: *The redshift distribution of the best dataset for a QuasarNET-like CNN model with regards to an independant test dataset.*

CNN.

Looking at the CNN predictions on the different templates presented in Figure 3.11, this logical result was found on Templates 1 to 3, for all redshifts.
However, a surprising result was observed for Template 0, which corresponds to the most complicated spectra to identify, as it lacks a Ly$\alpha$ line.

The result is shown in Figure 6.3. Looking at the blue curve, we can see that for Template 0, and for redshifts below 3, the CNN gains in precision as the magnitude increases, and therefore the noise decreases. This is a very surprising result, from which we can perhaps conclude that the CNN somehow learns a noise pattern, at least for spectra with low Ly-$\alpha$ emission and low redshifts.

## 6.3 Customized loss function

The loss function to be minimized, as defined in [7], is of the form

$$\mathcal{L} = C_{err} + R_{err}$$

With $C_{err}$ the error on the classification task, i.e. determining whether it is an LBG

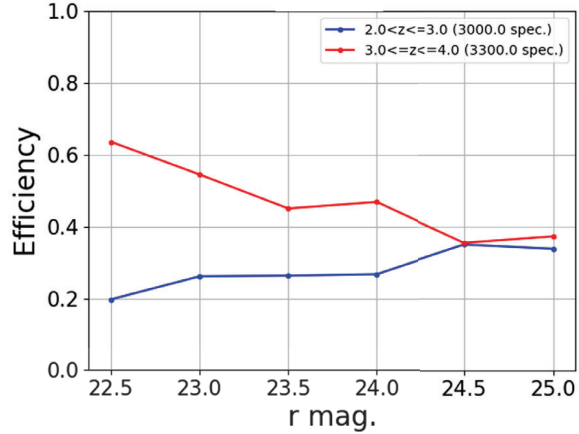

Figure 6.3: *The efficiency of the initial CNN on synthetic spectra as a function of the magnitude (r mag.) which is inversely proportional to the noise, for different redshift noises. The spectra on which the predictions are made are only Template 0 spectra, i.e without Ly-α emission line. The blue line represent the results on spectra with redshifts between 2.0 and 3.0 and the red one between 3.0 and 4.0*

or not, and $R_{err}$ the error on the regression task, i.e. determining the redshift of the galaxy, in the case of a LBG.

The idea is to introduce a gamma coefficient as :

$$\mathcal{L} = C_{err} + \gamma * R_{err}$$

to give more or less importance to one or other of the two tasks during gradient descent. Some initial tests have been carried out, and the results seem to vary according to the $\gamma$ parameter, but this idea needs to be explored further once the final model has been adopted, to try and find the right compromise between regression accuracy and classification efficiency.

# Chapter 7

# Conclusion and perspectives

## 7.1 Conclusion

The work carried out over 5 months was very interesting and rewarding, both in terms of cosmology and machine learning. Numerous ideas were tested, sometimes resulting in success and sometimes in failure.

The scope of this work as part of the DESI collaboration has been a great source of motivation and it will be very interesting to see the results of this collaboration, in the hope that it will help us to better understand the mysteries of the universe.

After introducing the context of the internship and the important concepts of cosmology in Chapter 1, the existing CNN model, QuasarNET, was presented in Chapter 2.2. The process of creating the best training dataset, presented in Chapter 3, enabled us to go from a dataset of around 3,000 spectra to one of over 60,000, with a clear gain in efficiency. Improvements were also achieved through modifications to the model, presented in Chapter 4, including the introduction of SNR and hyperparameter tuning. Finally, the results of the best models were presented in Chapter 5, and we can see that the objective of increasing the model's Efficiency-Purity trade-off has been achieved.

## 7.2 Perspectives

Even if this model has been improved, it is far from perfect. Throughout the internship, certain avenues could have been explored, or explored in greater detail, but due to lack of time, were not. It would be interesting to explore these avenues further as part of our work to improve the model.

- **Preprocessing.** Even if the spectra given as input to the CNN have already been transformed, in particular to set their average to 0 and by reducing their amplitude, it might be worth looking at whether other operations could facilitate the training of the CNN. The same is possible for the SNR.

- **Data augmentation.** It is still possible to imagine transformations to increase the size of the dataset, such as completely masking part of the spectrum. It would also be beneficial to create "online" transformations, i.e. ones that are randomly generated at each epoch, rather than creating upstream "offline" transformed spectra before training the model.

- **Transformers.** With the current boom in neural networks, it would be possible to imagine a completely different architecture, with the potential use of Transformers or other modern architectures, which are becoming increasingly efficient.

- **Hypertuning.** Hypertuning is a very time-consuming stage, and model convergence after a certain period of time is not necessarily optimal. It would probably be a good idea to run hypertuning again on fairly narrow ranges of possible hyperparameters values, initialized around the results of the hyperparameters tunings done in this project.

- **Additional data.** Having access to additional data would be an excellent asset to make the model more robust. New LBG readings could be beneficial for training the model. An additional human classification and regression of objects would also enable us to more reliably determine the redshift and nature of potential LBGs, and thus increase the size of the dataset.

# Appendix A

# Internship documents

## A.1 Gantt Diagram and assessment report

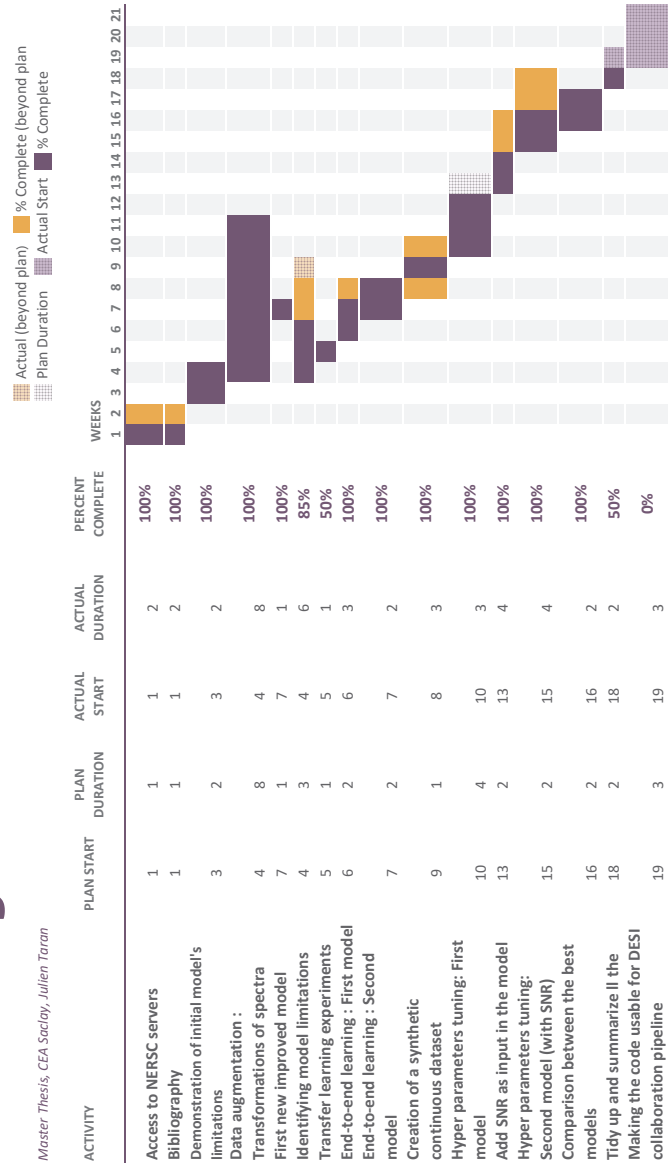

Figure A.1: *Gantt Diagram of the project*

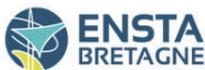

| Organisme | CEA Saclay |
|---|---|
| Dates du stage | 06/03/23 – 28/07/23 |
| NOM, Prénom du stagiaire | TARAN Julien |

| | F (échec) | E (insuffisant) | D (passable) | C (assez bien à bien) | B (bien à très bien) | A (remarquable) |
|---|---|---|---|---|---|---|
| **Critères d'intégration – Savoir être** | | | | | | |
| Adaptabilité | | | | | | x |
| Disponibilité | | | | | | x |
| Culture de l'entreprise | | | | | | x |
| Puissance de travail | | | | | | x |
| Qualité d'expression | | | | | | x |
| **Conduite du projet** | | | | | | |
| Identification des tâches | | | | | | x |
| Organisation/répartition des tâches dans le temps | | | | | | x |
| Respect des délais des livrables demandés | | | | | | x |
| Force de proposition | | | | | | x |
| Éventuellement : travail en équipe | | | | | | x |
| **Rapport de stage** | | | | | | |
| Forme (présentation, style…) | | | | | | x |
| Fond (exactitude) | | | | | | x |
| Exploitabilité par l'organisme | | | | | | x |
| **Appréciation de la formation ENSTA Bretagne** | | | | | | |
| Les compétences scientifiques et techniques répondent à mes attendus | | | | | | x |
| Les compétences méthodologiques répondent à mes attendus | | | | | | x |
| Sur quels sujets a-t-il fallu former le stagiaire avant qu'il ne soit autonome ? | Astronomie et cosmologie : essentiellement sur les spectres d'une galaxie dans le domaine visible et sur la mesure de son décalage vers le rouge | | | | | |
| Quelles seraient les compétences ou les contenus de formation à renforcer ? | Connaissances d'unix et utilisation de cluster de CPU/GPU | | | | | |

**Appréciation générale**
Julien a réalisé un très bon stage au sein du groupe de cosmologie du CEA à Saclay (Irfu/DPhP). Il a rapidement appris les notions d'astronomie nécessaires pour son travail. Il a su mettre en œuvre des techniques d'intelligence artificielle de reconnaissance d'images (CNN) pour mesurer le décalage vers le rouge de galaxies. Il a proposé des méthodes nouvelles et originales pour optimiser les performances de la méthode.
En outre, il a interagi avec des experts d'universités américaines ou de Thalés, travaillant sur le même sujet. Il a su pleinement tirer profit de ces discussions avec ces experts pour améliorer l'architecture du CNN.
**Si vous disposez d'un poste correspondant au profil du stagiaire, souhaiteriez-vous lui proposer ?  OUI**

NOM, Prénom du tuteur entreprise :  Christophe Yèche                                    Date : 05/07/23
Fonction : Directeur de Recherche, Chef du groupe de Cosmologie          Signature :

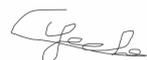

Figure A.2: *Assessment report of the internship*

# Appendix B

# LBG spectrum lines

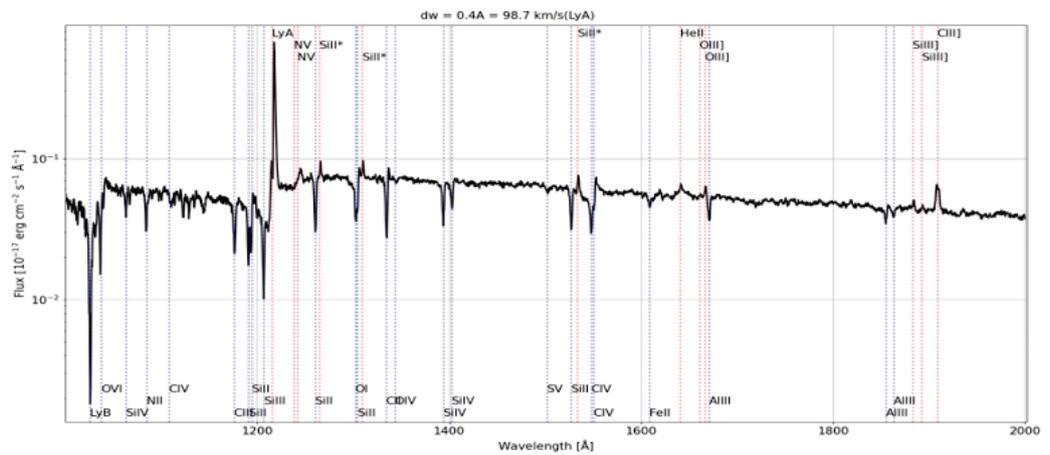

Figure B.1: *The absorption and emission lines of a LBG spectrum. Figure obtained by stacking several spectra. Credit : Christophe Magneville*

# Appendix C

# Test dataset and best training dataset

## C.1 Test dataset

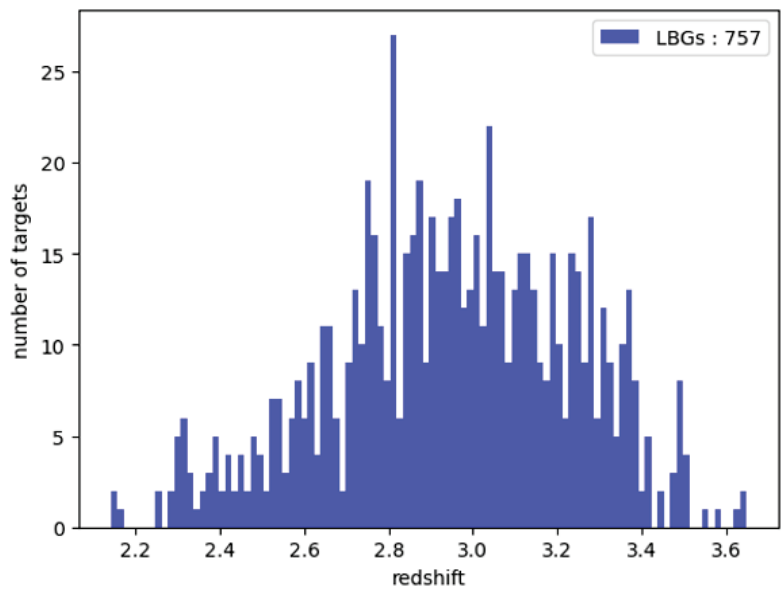

Figure C.1: *The test dataset of real spectra, used to evaluate models performances on classic redshifts around 3.*

## C.2 Creation of training dataset

Thanks to data augmentation presented in 3, a lot of sub-datasets have been created. With a combination of these sub-datasets, training datasets were created in order to maximize the performances of the model.

## C.3 Evaluation of the datasets

To determine the best training dataset, presented in Section 3.6, 16 combinations of datasets have been tried, and only the best ones kept. These datasets are combinations of sub-datasets presented in Appendix Section C.2 Finally, the dataset 1 got overall the best results.

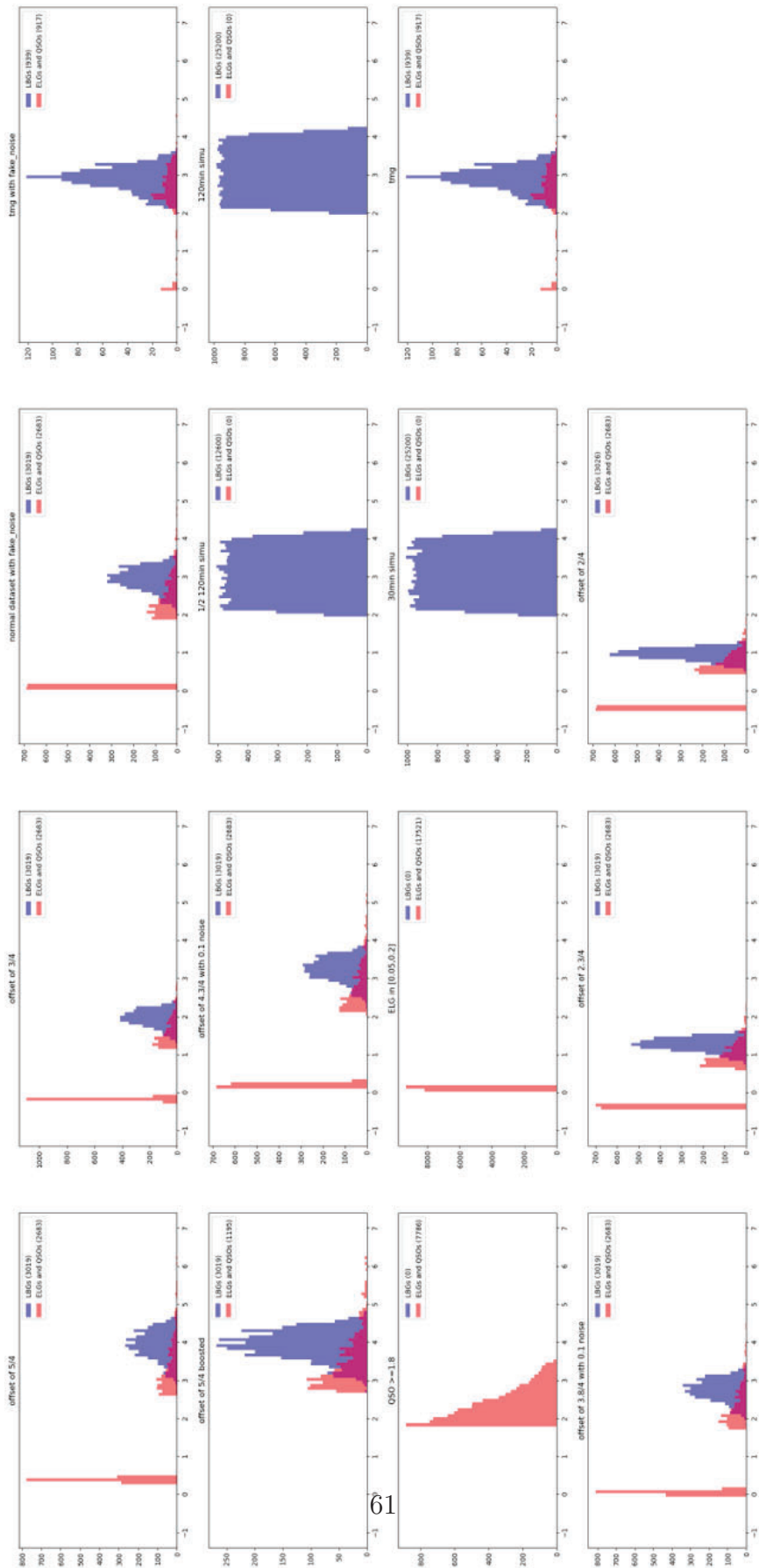

Figure C.2: *The distribution of LBGs and contaminants (ELG/QSO) in the sub-datasets*

```
data_to_train_1 = [
    'offset of 5/4',
    'offset of 3/4',
    'normal dataset with fake_noise',
    'tmg with fake_noise',
    'offset of 5/4 boosted',
    'offset of 4.3/4 with 0.1 noise',
    # '1/2 120min simu',
    '120min simu',
    'QSO >=1.8',
    'ELG in [0.05,0.2]',
    # '30min simu',
    # 'tmg',
    # 'offset of 3.8/4 with 0.1 noise',
    # 'offset of 2.3/4',
    # 'offset of 2/4',
    # '1/2 30min simu',
    # '1/3 30min simu'
    # '1/2 QSO >=1.8',
    # '1/2 ELG in [0.05,0.2]'
    # '1/4 QSO >=1.8',
    # '1/4 ELG in [0.05,0.2]'
],
```

```
data_to_train_2 = [
    'offset of 5/4',
    'offset of 3/4',
    'normal dataset with fake_noise',
    'tmg with fake_noise',
    'offset of 5/4 boosted',
    'offset of 4.3/4 with 0.1 noise',
    '1/2 120min simu',
    # '120min simu',
    'QSO >=1.8',
    'ELG in [0.05,0.2]',
    # '30min simu',
    # 'tmg',
    # 'offset of 3.8/4 with 0.1 noise',
    # 'offset of 2.3/4',
    # 'offset of 2/4',
    # '1/2 30min simu',
    '1/3 30min simu',
    # '1/2 QSO >=1.8',
    # '1/2 ELG in [0.05,0.2]',
    # '1/4 QSO >=1.8',
    # '1/4 ELG in [0.05,0.2]'
],
```

```
data_to_train_7 = [
    'offset of 5/4',
    'offset of 3/4',
    'normal dataset with fake_noise',
    'tmg with fake_noise',
    'offset of 5/4 boosted',
    'offset of 4.3/4 with 0.1 noise',
    # '1/2 120min simu',
    '120min simu',
    # 'QSO >=1.8',
    # 'ELG in [0.05,0.2]',
    # '30min simu',
    # 'tmg',
    # 'offset of 3.8/4 with 0.1 noise',
    # 'offset of 2.3/4',
    # 'offset of 2/4',
    # '1/2 30min simu',
    # '1/3 30min simu'
    '1/2 QSO >=1.8',
    '1/2 ELG in [0.05,0.2]',
    # '1/4 QSO >=1.8',
    # '1/4 ELG in [0.05,0.2]'
],
```

```
data_to_train_10 = [
    'offset of 5/4',
    'offset of 3/4',
    'normal dataset with fake_noise',
    'tmg with fake_noise',
    'offset of 5/4 boosted',
    'offset of 4.3/4 with 0.1 noise',
    # '1/2 120min simu',
    # '120min simu',
    # 'QSO >=1.8',
    # 'ELG in [0.05,0.2]',
    # '30min simu',
    # 'tmg',
    # 'offset of 3.8/4 with 0.1 noise',
    # 'offset of 2.3/4',
    # 'offset of 2/4',
    # '1/2 30min simu',
    # '1/3 30min simu'
    '1/2 QSO >=1.8',
    '1/2 ELG in [0.05,0.2]',
    # '1/4 QSO >=1.8',
    # '1/4 ELG in [0.05,0.2]'
],
```

```
data_to_train_13 = [
    'offset of 5/4',
    'offset of 3/4',
    'normal dataset with fake_noise',
    'tmg with fake_noise',
    'offset of 5/4 boosted',
    'offset of 4.3/4 with 0.1 noise',
    # '1/2 120min simu',
    '120min simu',
    # 'QSO >=1.8',
    # 'ELG in [0.05,0.2]',
    # '30min simu',
    # 'tmg',
    # 'offset of 3.8/4 with 0.1 noise',
    # 'offset of 2.3/4',
    # 'offset of 2/4',
    # '1/2 30min simu',
    # '1/3 30min simu'
    # '1/2 QSO >=1.8',
    # '1/2 ELG in [0.05,0.2]',
    '1/4 QSO >=1.8',
    '1/4 ELG in [0.05,0.2]'
],
```

Figure C.3: *The sub-datasets composition of the best five models. "data_to_train_X" corresponds to "dataset X" in the following section and one can see the distribution of the data for each sub-dataset in C.2*

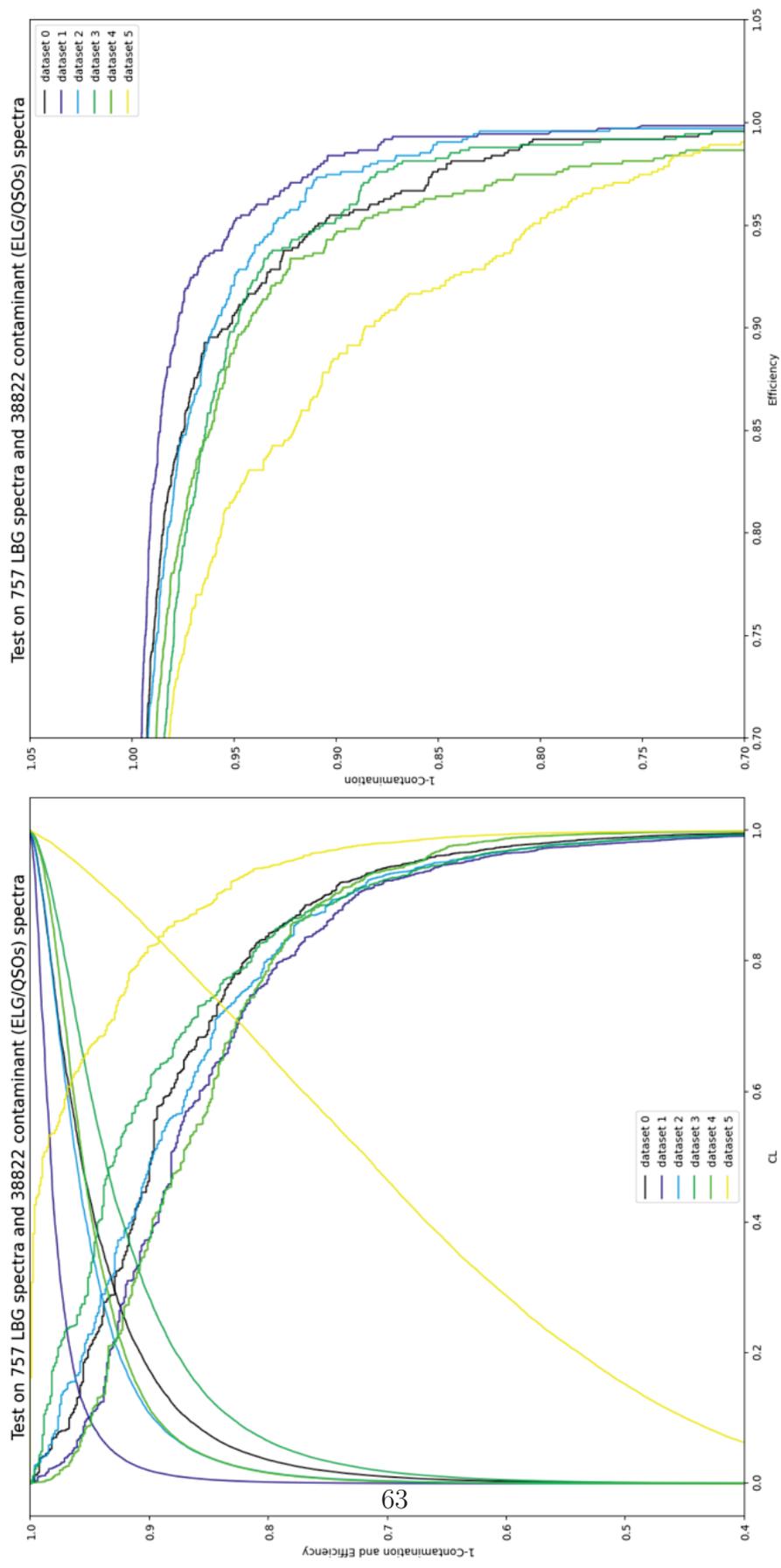

Figure C.4: *Comparison between the six first datasets*

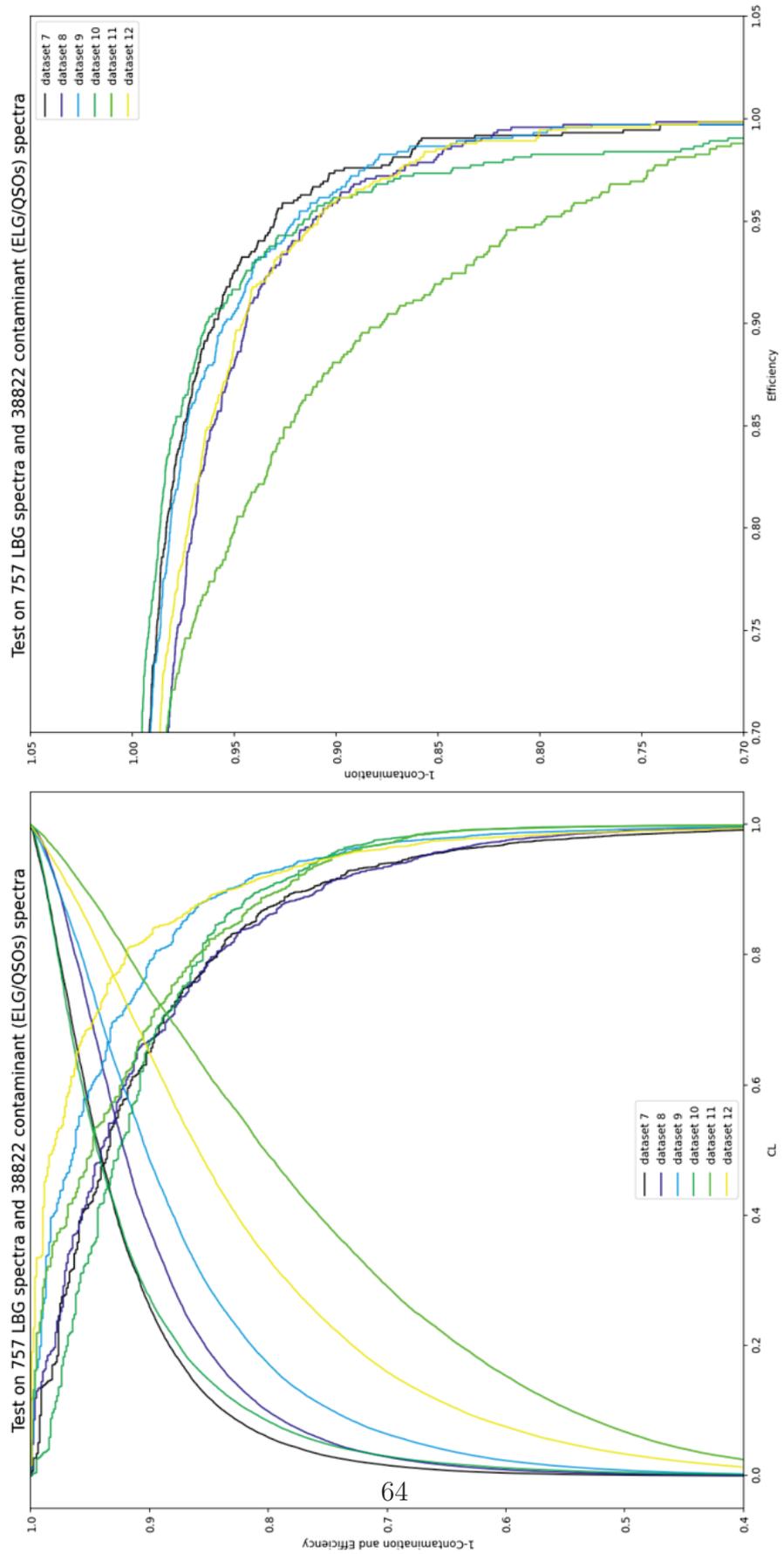

Figure C.5: *Comparison between datasets seven to twelve.*

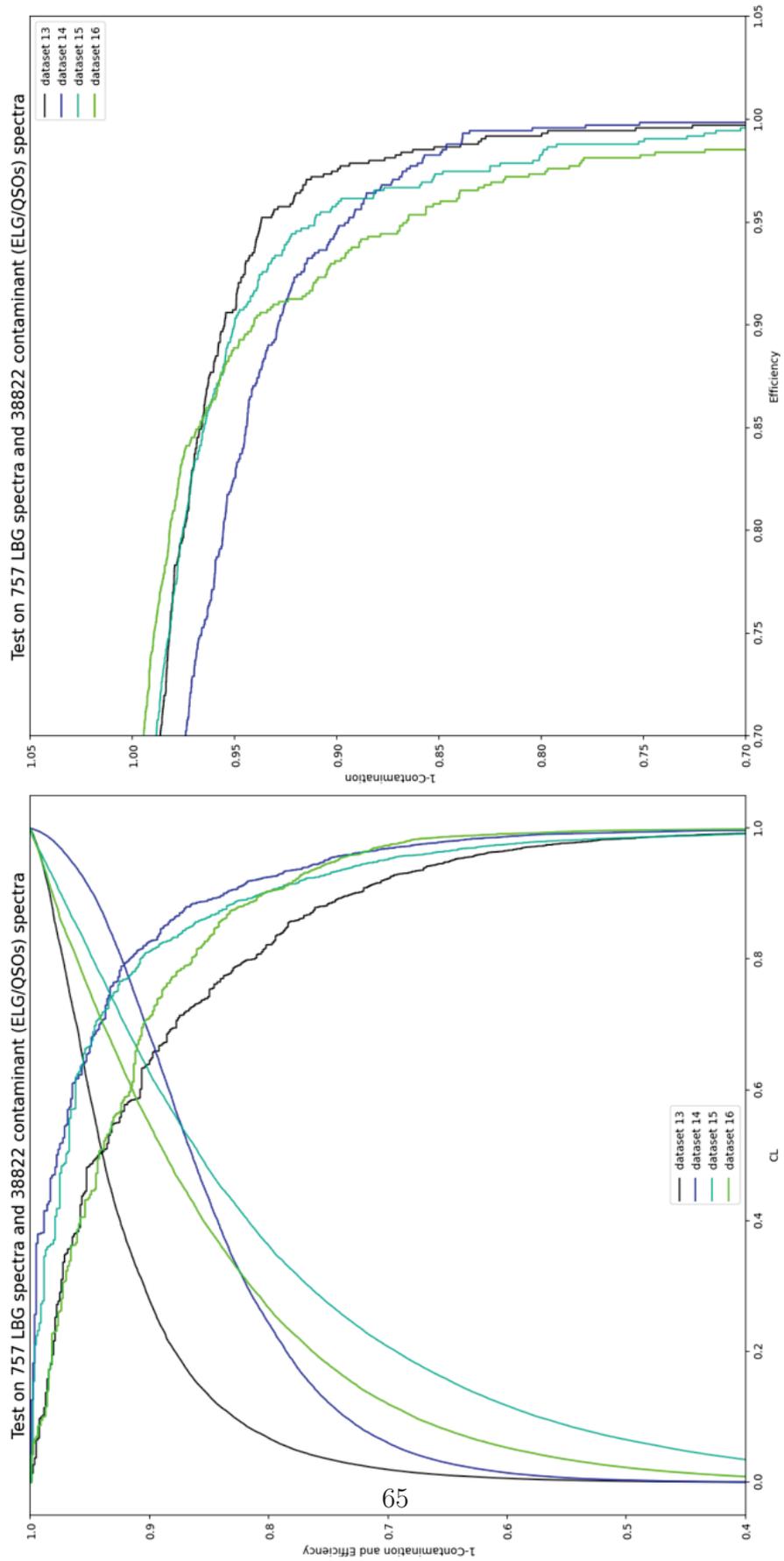

Figure C.6: *Comparison between the four last datasets.*

# Appendix D

# CNN Architecture

## D.1 Details on QuasarNET architecture

Batch normalization is computed differently during the training and the testing phase.

### B.1.1) Training

At each hidden layer, Batch Normalization transforms the signal as follow :

$$(1)\ \mu = \frac{1}{n}\sum_i Z^{(i)} \qquad (2)\ \sigma^2 = \frac{1}{n}\sum_i (Z^{(i)} - \mu)^2$$

$$(3)\ Z^{(i)}_{norm} = \frac{Z^{(i)} - \mu}{\sqrt{\sigma^2 - \epsilon}} \qquad (4)\ \breve{Z} = \gamma * Z^{(i)}_{norm} + \beta$$

The BN layer first determines the **mean μ** and **the variance σ²** of the activation values across the batch, using (1) and (2).

It then **normalizes the activation vector Z^(i)** with (3). That way, each neuron's output follows a standard normal distribution across the batch. (ε is a constant used for numerical stability)

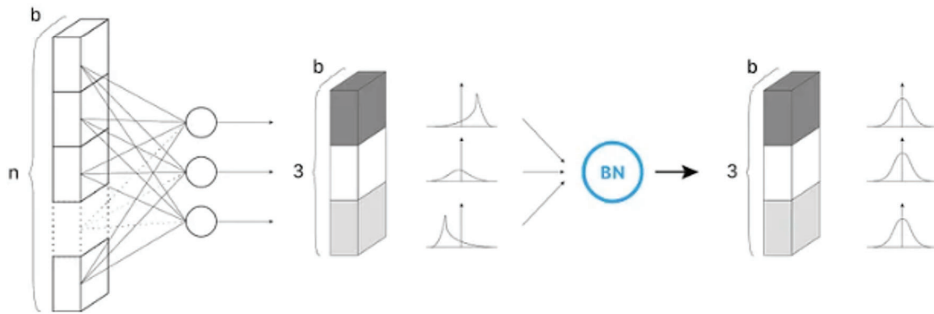

Batch Normalization first step. Example of a 3-neurons hidden layer, with a batch of size b. Each neuron follows a standard normal distribution. | Credit : author - Design : Lou HD

At each iteration, the network computes the mean μ and the standard deviation σ corresponding to the current batch. Then it trains γ and β *through* gradient descent, using an Exponential Moving Average (EMA) to give more importance to the latest iterations.

## D.2  End-To-End learning results

### B.1.2) Evaluation

Unlike the training phase, **we may not have a full batch to feed into the model during the evaluation phase.**

To tackle this issue, we compute ($\mu$_pop , $\sigma$_pop), such as :

- $\mu$_pop : estimated mean of the studied population ;
- $\sigma$_pop : estimated standard-deviation of the studied population.

Those values are computed using all the ($\mu$_batch , $\sigma$_batch) determined during training, and directly fed into equation (3) during evaluation (instead of calling (1) and (2)).

*Remark : We will discuss that matter in depth in section C.2.3 : "Normalization during evaluation".*

It finally calculates the **layer's output $\hat{Z}$ (i)** by applying a linear transformation with $\gamma$ and $\beta$, two trainable parameters (4). Such step allows the model to choose the optimum distribution for each hidden layers, by adjusting those two parameters :

- $\gamma$ allows to adjust the standard deviation ;
- $\beta$ allows to adjust the bias, shifting the curve on the right or on the left side.

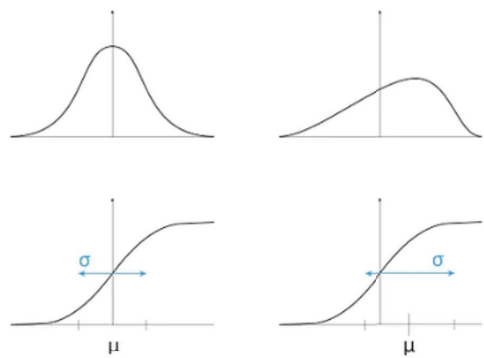

Benefits of $\gamma$ and $\beta$ parameters. Modifying the distribution (on the top) allows us to use different regimes of the nonlinear functions (on the bottom). | Credit : author — Design : Lou HD

Figure D.1: *The Batch Normalization principle. From [10]*

### 3. ReLU (Rectified Linear Unit) Activation Function

The ReLU is the most used activation function in the world right now.Since, it is used in almost all the convolutional neural networks or deep learning.

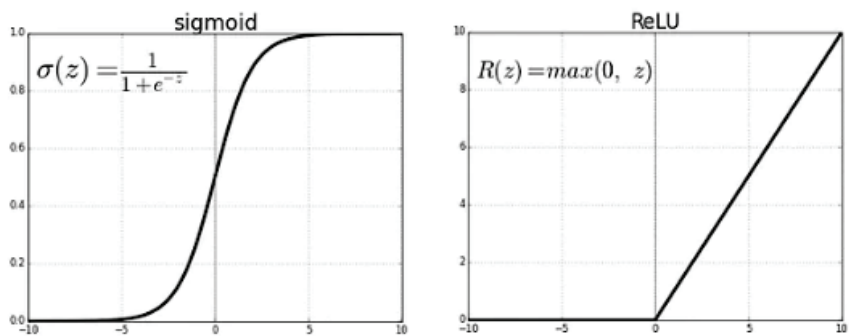

Fig: ReLU v/s Logistic Sigmoid

As you can see, the ReLU is half rectified (from bottom). f(z) is zero when z is less than zero and f(z) is equal to z when z is above or equal to zero.

**Range:** [ 0 to infinity)

The function and its derivative **both are monotonic.**

But the issue is that all the negative values become zero immediately which decreases the ability of the model to fit or train from the data properly. That means any negative input given to the ReLU activation function turns the value into zero immediately in the graph, which in turns affects the resulting graph by not mapping the negative values appropriately.

Figure D.2: *The ReLu activation function principle. From [12]*

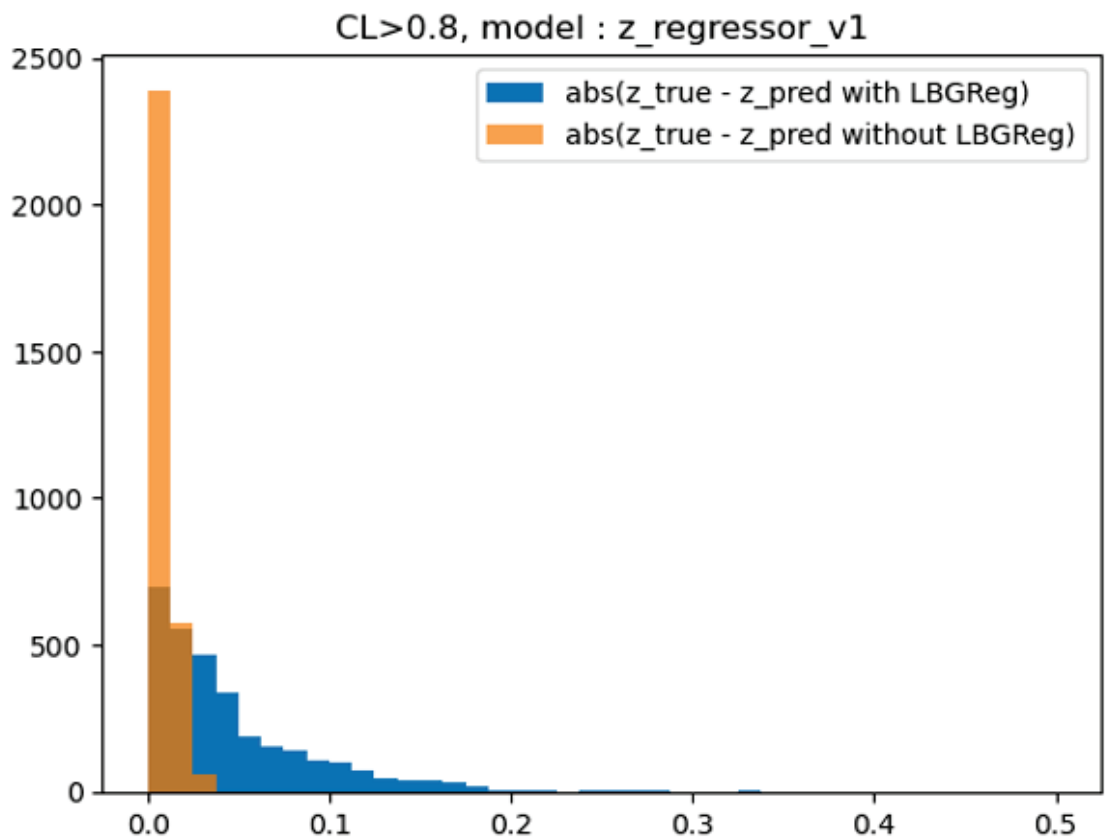

Figure D.3: *Comparison between the classis QuasarNET model architecture and the **first** "End-to-End" type solution presented in Section 4.2. named LBGReg. The x axis is the absolute difference between the redshift predicted and the real redshift of LBGs, and the y axis represents the number of targets.*

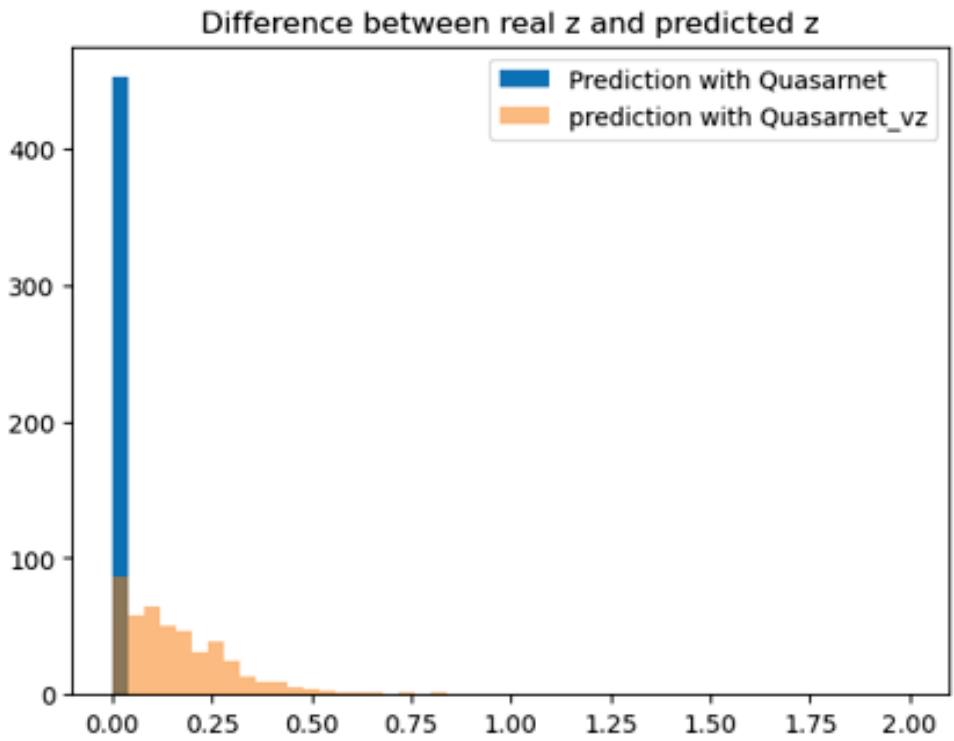

Figure D.4: *Comparison between the classis QuasarNET model architecture and the **second** "End-to-End" type solution presented in Section 4.2. named Quasarnet_vz. The x axis is the absolute difference between the redshift predicted and the real redshift of LBGs, and the y axis represents the number of targets.*